\documentclass[journal]{IEEEtran}

\usepackage{graphicx}
\usepackage{amssymb}
\usepackage{amsmath}

\usepackage{algorithm}
\usepackage{algorithmic}
\usepackage{enumerate}
\usepackage{threeparttable}
\usepackage{multirow}
\usepackage{bm}
\usepackage{soul, color, xcolor}

\usepackage[colorlinks]{hyperref}
\begin{document}

% \author{Zhifei Li,~\IEEEmembership{Student Member, IEEE}
% 	and Zhaoli Zhang,~\IEEEmembership{Member, IEEE}
	
% \IEEEcompsocitemizethanks{\IEEEcompsocthanksitem Z. Li and Z. Zhang are with the National Engineering Research
% 		Center for E-Learning, Central China Normal University, Wuhan 430079,
% 		China. \\E-mail: zhifei1993@gmail.com; 
% 	}

% } 

\author{{Ke Shi\textsuperscript{1}}, {Yan Zhang\textsuperscript{1}}, {Miao Zhang}, {Lifan Chen}, {Jiali Yi}, {Kui Xiao}, {Xiaoju Hou}, and {Zhifei Li\textsuperscript{*}}
	
\IEEEcompsocitemizethanks{
\IEEEcompsocthanksitem This work was supported in part by the National Natural Science Foundation of China (No. 62207011, 62377009, 62407013), the Natural Science Foundation of Hubei Province of China (No. 2025AFB653). K. Shi and Y. Zhang are co-first authors. (Corresponding author: Zhifei Li)
\IEEEcompsocthanksitem K. Shi, Y. Zhang, M. Zhang, L. Chen, J. Yi, K. Xiao, and Z. Li are with the School of Computer Science, Hubei University, Wuhan 430062, China, Hubei Key Laboratory of Big Data Intelligent Analysis and Application, Hubei University, Wuhan 430062, China, and also with the Key Laboratory of Intelligent Sensing System and Security (Hubei University), Ministry of Education, Wuhan 430062, China. X. Hou is with the Institute of Vocational Education, Guangdong Industry Polytechnic University, Guangzhou 510300, China. 
 }

} 

\title{Structurally Refined Graph Transformer for Multimodal Recommendation}

\maketitle

\markboth{IEEE Transactions on Multimedia}
{\MakeLowercase{\textit{Shi et al.}}: SRGFormer}

% \definecolor{lb}{rgb}{1.0, 0.835, 0.835}
% \sethlcolor{lb}

% abstract or keywords.
\begin{abstract}
Multimodal recommendation systems utilize various types of information, including images and text, to enhance the effectiveness of recommendations. The key challenge is predicting user purchasing behavior from the available data. Current recommendation models prioritize extracting multimodal information neglecting the distinction between redundant and valuable data. They also rely heavily on a single semantic framework (e.g., local or global semantics), resulting in an incomplete or biased representation of user preferences, particularly those less expressed in prior interactions. Furthermore, these approaches fail to capture the complex interactions between users and items limiting the model’s ability to meet diverse users. To address these challenges, we present SRGFormer, a structurally optimized multimodal recommendation model. By modifying the transformer for better integration into our model, we capture the overall behavior patterns of users. Then, we enhance structural information by embedding multimodal information into a hypergraph structure to aid in learning the local structures between users and items.
Meanwhile, applying self-supervised tasks to user-item collaborative signals enhances the integration of multimodal information, thereby revealing the representational features inherent to the data's modality. Extensive experiments on three public datasets reveal that SRGFormer surpasses previous benchmark models, achieving an average performance improvement of 4.47\% on the Sports dataset. Our code is available at \url{https://github.com/HubuKG/SRGFormer}.
\end{abstract}


\section{Introduction}
The swift growth of online data has led platforms to implement multimodal recommendation systems, initially using collaborative filtering (CF) to analyze user preferences from historical interactions \cite{1,2}. However, CF struggles to handle sparse or non-existent interaction records leading to less accurate predictions. Problems such as the cold start issue and sparse interaction data remain challenging to solve \cite{3}. To address these challenges, many studies have explored augmenting collaborative signals with additional information to learn user preferences better. Incorporating text and image data, user comments, and other modal information has become common practice. In contemporary research, multimodal recommendation systems have emerged as a new alternative to traditional recommendation methods \cite{51,52}. They improve the capture of user behavior patterns by strengthening traditional user-item interaction models and incorporating multiple information modalities. For instance, MMGCN constructs user-item bipartite graphs for specific modalities and uses graph neural networks to learn fine-grained user and item feature representations across multimodal such as vision, sound, and text \cite{28}. MMKGAT leverages attention mechanisms in knowledge graphs to refine item representations \cite{6}.

\begin{figure}
    \centering    
    \includegraphics[width=0.95\linewidth,height=0.95\linewidth]{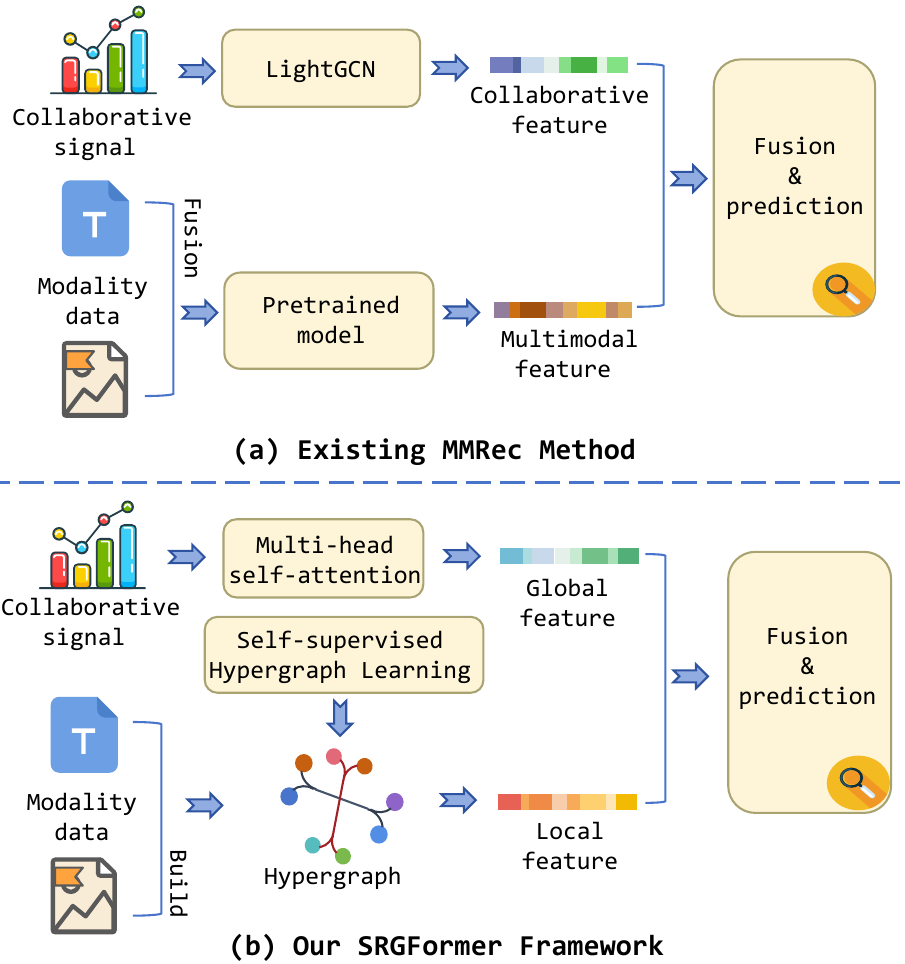}
    \caption{Intuitions of existing MMRec methods and SGFormer.}
    \label{fig:1}
\end{figure}

Although these methods are effective, they still have limitations, as most of them primarily focus on users' historical interactions while ignoring the structural features between different pieces of information \cite{7,8}. Users have unique preferences, and an individual user may demonstrate varying levels of interest across projects \cite{9,10,11}. The difference primarily stems from the structural interaction between the product characteristics and the user's buying intention. Hence, learning structural information is considered essential in recommendation systems. Fig. \ref{fig:1}(a) illustrates the general framework of a traditional multimodal recommendation model that employs lightweight neural networks and pre-trained models (e.g., BERT, ViT \cite{14}) to extract interaction information and multimodal features \cite{12, 13}. In contrast, Fig. \ref{fig:1}(b) demonstrates that our proposed framework employs customized algorithmic techniques to effectively capture the fine-grained node relationships and optimize various information structures. Compared to traditional models, our approach refines user preferences across multiple dimensions and emphasizes the semantic relationships between nodes. By leveraging the advantages of hypergraphs and multi-head attention mechanisms to analyze graph structural features, redundant and valuable information can be effectively filtered and analyzed \cite{53,54}. Traditional hypergraphs often rely on large amounts of manually labeled data \cite{21}. Consequently, we introduce self-supervised learning through graph structure reconstruction tasks, enabling the model to predict missing parts autonomously. This strategy allows the model to more effectively and dynamically analyze the sensitivity and structural relationships between users and items, enhancing the accuracy of user behavior predictions \cite{15}.

In this paper, we aim to accurately capture user behavior patterns and make predictions by combining the structural connections between users and items and applying self-supervised tasks to multimodal information as an aid. The SRGFormer framework, a self-supervised model based on Transformer, is introduced and consists of three core modules: multimodal interaction and modeling, structural information interaction and modeling, and fusion and prediction. Initially, we use a transformer to assign attention weights to items based on each user's historical interactions learning global behavior patterns and revealing overall structure. Then, by constructing a hypergraph based on multimodal information, the relevant local and global structures are integrated to strengthen the user interaction framework. We also extract various modal information from the modal extractor and assign self-supervised tasks to capture interactions between Different modal types of data. Finally, the refined user structure is fused with user preferences across different modalities to derive the final preference score, enhancing recommendation accuracy. Our primary contributions can be summarized as follows:

\begin{itemize}
    \item We propose the SRGFormer model which emphasizes the impact of different items on the user-item semantic structure and the refined user preference structure. Refining user preferences through global and local structures allows the model to capture user preferences more accurately.
    \item We assign self-supervised tasks to different modalities to better learn their interactions helping the model further explore modal information and improve prediction accuracy and model generalization.
    \item We conduct comprehensive experiments on three distinct public datasets. In terms of automatic evaluation metrics Recall and NDCG, SRGFormer achieves an average improvement of 4.46\% on the Baby dataset and 4.47\% on the Sports dataset, compared to the baseline model.
\end{itemize}

The subsequent sections are organized as follows: Section \ref{section2} presents a comprehensive discussion of the relevant literature. The intricacies of our model are expounded in Section \ref{section3}. Experimental results and corresponding analyses are presented in Section \ref{section4}. Finally, Section \ref{section5} offers concluding remarks providing closure to this paper.

\section{Related work}
\label{section2}

Recommender systems represent a significant application of artificial intelligence technology, particularly in online shopping \cite{16,17}. These systems utilize powerful databases to identify and analyze all products subsequently recommending items that users may potentially be interested in \cite{18}. To enhance the accuracy of recommendations, incorporating different modal information or restructuring the data has become an important approach to addressing the cold start and data sparsity problems \cite{19,20}.

\subsection{Multimodal Recommendation}
Traditional recommendation methods primarily rely on learning from user-item interactions to generate accurate recommendations \cite{22}. However, this approach has inherent limitations. Most datasets face data issues such as cold start \cite{20} and data sparsity \cite{21} hindering traditional recommendation systems from effectively capturing users' complex preferences in these scenarios \cite{23}.

To tackle the issues of data sparsity and cold start, recommendation systems have started integrating multimodal information. For instance, DualGNN innovates in modeling multimodal user preferences for micro-video recommendations, aiming to address the limitations of existing methods that fuse user preferences from different modalities in a unified manner \cite{24}. It also tackles the common issue of missing modalities in micro-video recommendations. VBPR incorporates visual information using a pre-trained deep CNN to extract image features, thereby enhancing the learning of users' features from their historically interacted items \cite{25}. AMR identifies vulnerabilities in current multimedia recommendation systems such as VBPR and fortifies them using adversarial learning \cite{24}. VMCF builds a product affinity network where products serve as nodes and view relationships act as edges leveraging both visible and latent relational information. Simultaneously, it utilizes Bayesian Personalized Ranking to make recommendations more accurate \cite{26,27}. Furthermore, techniques from self-supervised learning, such as those implemented in MMSSL, effectively enhance user-item relationship exploration by utilizing self-supervised learning within a multi-modal framework. MMSSL leverages self-supervised signals to learn users' modal-aware preferences and cross-modal dependencies addressing the limitations of label dependency and sparse user behavior data \cite{28}. By employing modal-aware interactive structure learning and cross-modal contrastive learning, it enhances data representation and model robustness, thereby capturing users' complex preferences more accurately \cite{29}. Building on this line of research, recent studies emphasize robustness and efficiency in multimodal recommendation. Mirror Gradient \cite{49} guides model parameters toward flat minima via alternating gradient updates to enhance robustness against noise and distribution shifts, while IISAN \cite{48} employs a decoupled fine-tuning framework with intra- and inter-modal side adapted networks to maintain accuracy and substantially reduce GPU memory and training time.

\subsection{Attention mechanism for MMRec}
Attention mechanisms are inspired by human cognitive processes mimicking how individuals allocate varying levels of attention to different aspects of various items. This approach dynamically learns user preferences assigning distinct sensitivities to the same user's interactions with different items \cite{30,31,32}.

Traditional multimodal recommendation systems often overlook users' varying preferences for different items, thereby limiting the predictive performance of the models. However, reinforcement learning for users can be enhanced by attention weight distribution. For example, GFormer enhances data representation by integrating generative self-supervised learning with a graph transformer architecture \cite{33}.  MKGformer achieves fine-grained fusion through shared $Q, K, V$ parameters and a context-aware fusion module \cite{34}. MGAT employs a gated attention mechanism to focus on users' local preferences \cite{35}. MARIO predicts user preferences by evaluating the impact of each modal on each interaction \cite{36}. FREEDOM uses an item-item graph for each modal \cite{37}, similar to LATTICE \cite{38}, but it freezes the graph before training and introduces degree-sensitive edge pruning to denoise the user-item interaction graph. Despite these advances, existing methods have yet to thoroughly evaluate the impact of individual interactions, which limits the potential of historical interaction data. Additionally, there is a need for a deeper exploration of attention mechanisms.

\begin{figure*}
    \centering
    \includegraphics[width=1\linewidth, height=0.5\linewidth]{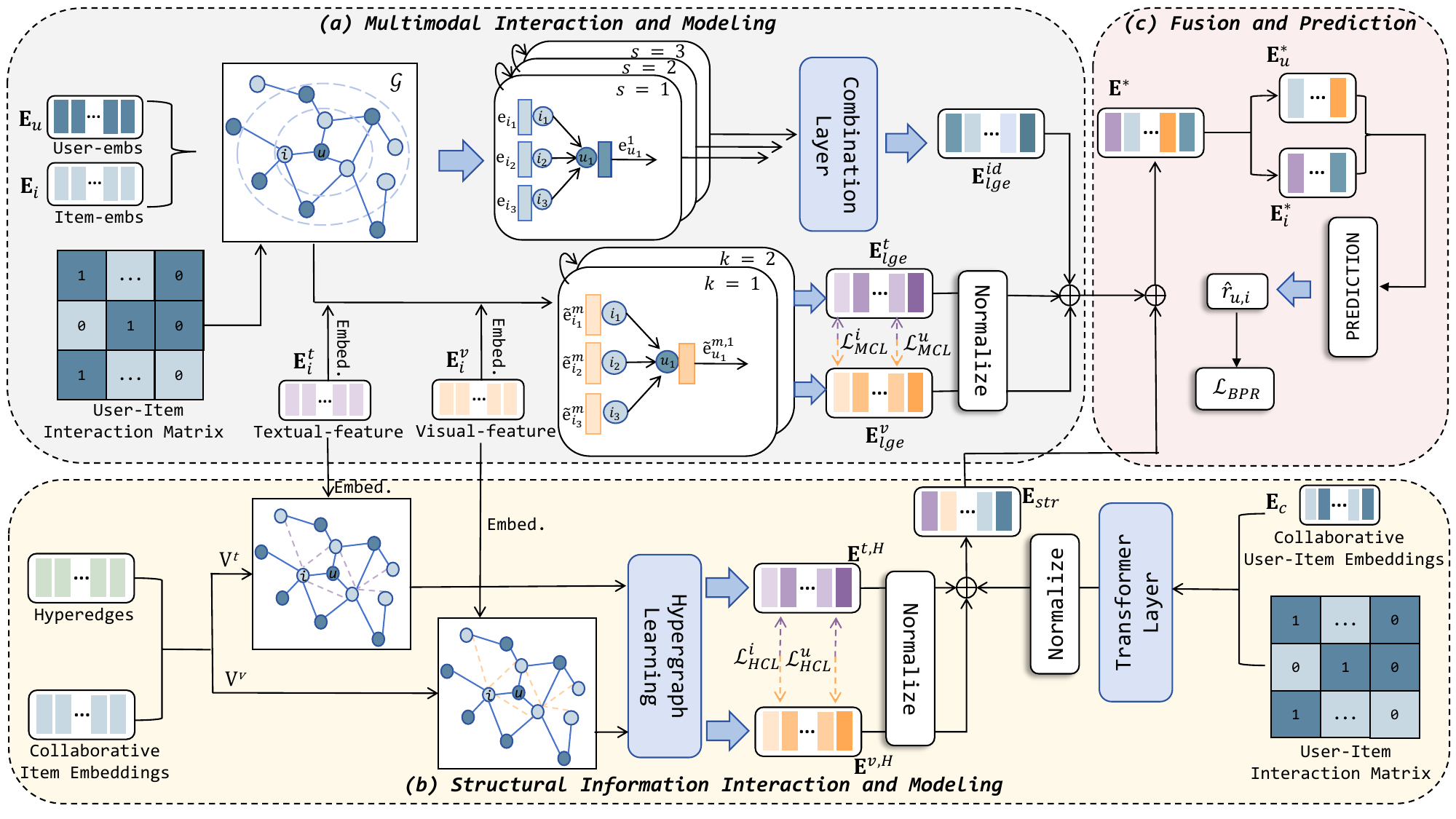}
    \caption{(a) The multimodal interaction and modeling module captures user item representations alongside semantic representations of modal information. (b) The structural information interaction and modeling module enhances the user's structural comprehension. (c) The Fusion and Prediction module amalgamates semantic information from various modalities, collaborative signals, and user structures to forecast user preference scores.}
    \label{fig:2}
\end{figure*}

\section{Methodology}
\label{section3}

This section provides a detailed exposition of SRGFormer. As depicted in Fig. \ref{fig:2}, SRGFormer features a comprehensive architecture consisting of three primary modules: (1) The multimodal interaction and modeling module: extract and enhance multimodal information. (2) The structural information interaction and modeling module: learn the refined structure of user needs. (3) The fusion and prediction module: conduct predictions and scoring of user preferences.

\subsection{Preliminaries}
We set user set as ${\cal U}=\{u\}$ and item set as ${\mathcal{I}}=\{i\}$. The ID embeddings of each user $u\in {\cal U}$ and each item $i\in {\mathcal{I}}$ are denoted as $\mathbf{e}_{u}\in\mathbb{R}^{d}$ and $\mathbf{e}_{i}\in\mathbb{R}^{d}$, respectively, where $d$ is the embedding dimension. The user-item interactions can be represented as a matrix ${\mathbf{R}}\in\mathbb{R}^{\left|{\cal U}\right|\times\left|\mathcal{I}\right|}$, in which the element $r_{u, i}$ within the matrix is assigned a value of one to indicate the presence of an interaction between user $u$ and item $i$;  conversely, it is set to zero in the absence of such interaction. In our system, a specific feature vector uniquely identifies and describes each user and item. Additionally, key attributes of each item, including descriptions, prices, brands, and images, are represented through both textual and visual feature vectors. Table \ref{table1} provides explanations for the various symbols used in the model.

% \usepackage{graphicx}
\begin{table}[]
\caption{Notations and descriptions.}
\renewcommand{\arraystretch}{1.1}
\resizebox{0.5\textwidth}{!}{%
\begin{tabular}{l|l}
\hline
Notations  & Explanations                        \\ \hline
$u , i$      & The set of users and items          \\
$\mathbf{E}_i , \mathbf{E}_u$  & The feature matrix of item and user \\
$\widetilde{\mathbf{E}}^{m, K}$                          & The last \textit{K}-th layer as modal-related embeddings \\
$\widetilde{\mathbf{e}}^m_i$ , $\widetilde{\mathbf{e}}^m_u$ & Transformed modal feature of item and user             \\
$\hat{r}_{u,i}$ & Final forecast score                \\
$\mathbb{R} $        & The real number set       \\
$\mathbf{H}_i^m , \mathbf{H}_u^m$    & Hyperedge dependency matrices of items and users       \\
$\mathbf{W}_m$       & Modality-related weight matrix      \\ \hline
\end{tabular}%
}
\label{table1}
\end{table}

\subsection{Multimodal Interaction and Modeling}
The Multimodal Interaction and Modeling Module aims to utilize modal weight matrices and LightGCN to extract multimodal features, thereby mapping the modal information into user and item embeddings \cite{11}. The embeddings of users and items are learned through multi-layer LightGCN message propagation on a user-item interaction graph formed by ID embeddings, thus capturing high-order connectivity. Specifically, the collaborative graph propagation function $\text{CGPROG}(\cdot)$ in $({\mathrm{s}}+1)$-th layer can be formatted as:
\begin{align}
\mathbf{E}^{s+1}=\text{CGPROG}(\mathbf{E}^{s})=(\mathbf{D}^{-\frac{1}{2}}\mathbf{AD}^{-\frac{1}{2}})\mathbf{E}^{s},
\end{align}
here, $\text{CGPROG}(\cdot)$ employs a simplified graph convolutional network method to efficiently propagate user and item embeddings across multiple interaction layers enhancing the feature representation of each modality. In this context, $\mathbf{A}$ represents an adjacency matrix in the embedding space $\mathbb{R}^{(|{\cal U}| + |\mathcal{I}|) \times (|{\cal U}| + |\mathcal{I}|)}$, constructed from the interaction matrix $\mathbf{R}$ which details the interactions between users $u$ and items $i$. A diagonal matrix $\mathbf{D}$ for $\mathbf{A}$ has elements $\mathbf{D}_{j,j}$ representing the count of non-zero entries in the \textit{j}-th row of $\mathbf{A}$. Initially, the embeddings matrix is set as $\mathbf{E}^{0} = \mathbf{E}^{id}$, where $\mathbf{E}^{id}$ are the ID embeddings. Subsequent layers build upon this initial mapping by using collaborative graph propagation to deepen connectivity and feature integration across different modalities. In this process, layer combination techniques are adopted to integrate all embeddings from the hidden layers,
\begin{align}
\mathbf{E}_{lge}^{id} = \text{LAYERCOMB}(\mathbf{E}^{0}, \mathbf{E}^{1}, \mathbf{E}^{2}, \ldots, \mathbf{E}^{S}),
\end{align}
where $\mathbf{E}_{lge}^{id} \in \mathbb{R}^{(|{\cal U}| + |\mathcal{I}|) \times d}$ represents the collaborative signals of users and items incorporating original neighbor information. The mean function is employed to implement $\text{LAYERCOMB}(\cdot)$ for embedding integration.

Based on the initial user-item collaboration signals extracted, we learn the influence of different modal features on users and items by mapping the features of these modal features into collaborative embeddings. The original item modal features are typically generated by pre-trained models such as BERT resulting in them existing in different vector spaces and having various dimensions. Therefore, before mapping them into user and item embeddings, the distinct modal features are uniformly projected into the embedding space $\mathbb{R}^{d}$ using pre-learned modal weight matrices:
\begin{align}
\tilde{\mathbf{e}}_i^m = \text{TRANSFORM}(\mathbf{e}_i^m) = \mathbf{e}_i^m \cdot \mathbf{W}_m,
\end{align}
here, $\tilde{\mathbf{e}}_i^m$ represents the mapped modal embeddings and $\text{TRANSFORM}(\cdot)$ denotes a projection function parameterized by pre-learned modal weight matrices $\mathbf{W}_m \in \mathbb{R}^{d_m \times d}$. When user ID embeddings are directly used as initial vectors for modal feature extraction, the embedded features tend to show coupling after matrix transformation resulting in unclear user preference relationships. Therefore, we initialize user modal features by using multi-layer embeddings and aggregating item modal features,
\begin{align}
\mathbf{e}_{u}^{m} = \frac{1}{|\mathcal{N}_u|} \sum_{i \in \mathcal{N}_u} \mathbf{e}_{i}^{m},
\end{align}
where $\mathcal{N}_u$ denotes the set of neighbors for user $u \in \mathcal{U}$ in the user-item interaction graph $\mathcal{G}$. This complete separation of ID embeddings from modal features helps avoid coupling to a certain degree. Thereafter, we can construct the modal feature matrix $\bm{\mathbf{E}}^m = \left[ \tilde{\mathbf{e}}_{u_1}^m, \ldots, \tilde{\mathbf{e}}_{u_{|\cal U|}}^m, \ldots, \tilde{\mathbf{e}}_{i_1}^m, \ldots, \tilde{\mathbf{e}}_{i_{|{\mathcal{I}}|}}^m \right]$ as initial input $\tilde{\mathbf{E}}^{m,0}$ to learn modal-related embeddings via implementing a light graph propagation function $\text{MGPROG}(\cdot)$,
\begin{align}
\widetilde{\mathbf{E}}^{m, k+1} = \text{MGPROG}(\widetilde{\mathbf{E}}^{m, k}) = (\mathbf{D}^{-\frac{1}{2}} \mathbf{A} \mathbf{D}^{-\frac{1}{2}}) \widetilde{\mathbf{E}}^{m, k}.
\end{align}

To further promote the integration between modalities and improve the learning of their interrelationships, corresponding self-supervised tasks are specified during the fusion process. Using the collaborative embeddings of users from different modalities as initial values, we define the loss for this process as follows:
\begin{align}
\mathcal{L}^{u}_{MCL} = \sum_{u \in \mathcal{U}} - \log \frac{\exp(s(\mathbf{E}^{v}_{u, lge}, \mathbf{E}^{t}_{u, lge}) / \tau)}{\sum_{u' \in \mathcal{U}} \exp(s(\mathbf{E}^{v}_{u', lge}, \mathbf{E}^{t}_{u', lge}) / \tau)},
\end{align}
here, $\mathbf{E}_{u,lge}^v$ and $\mathbf{E}_{u,lge}^t$ represent the embeddings of user $u$ in different modalities \textit{v} and \textit{t}, respectively. The function $s(\cdot)$ is cosine similarity. 

Ultimately, the higher-order modal embedding $\tilde{\mathbf{E}}^{m, K}$ is selected from the final $K$ layers as the embedding vector for the corresponding modality $m$.

\subsection{Structural Information Interaction}
For general data, the Multimodal Interaction and Modeling module usually suffices for extraction and learning. However, due to sparse and noisy information, two components are proposed in the Structural Information Interaction module: (1) Hypergraph Constructing, which enables hypergraph learning of different user modal local structures, and (2) Transformer Layer Constructing, which facilitates multi-head neural attention learning of users’ global structures. By integrating these components, we can refine the structures of all users ensuring a more accurate and robust learning process.

\subsubsection{Hypergraph Constructing}
Due to the incomplete and high-dimensional nature of explicit attribute information for most items, learnable implicit attribute vectors $\left\{ \mathbf{v}_a^m \right\}_{a=1}^A \left( \mathbf{v}_a^m \in \mathbb{R}^{d_m} \right)$ are defined as hyperedge embeddings in modality $m$. This involves assigning $A$ hyperedges to each item/user, thereby adaptively learning the dependency relationships between different users/items and implicit data. The following method is primarily used to construct hyperedge dependencies for users/items in a low-dimensional space:
\begin{align}
\mathbf{H}_i^m = \mathbf{E}_i^m \cdot \mathbf{V}^{m^\top}, \quad \mathbf{H}_u^m = \mathbf{A}_u \cdot \mathbf{H}_i^{m^\top},
\end{align}
among them, $\mathbf{H}_i^m \in \mathbb{R}^{|\mathcal{I}| \times A}$ and $\mathbf{H}_u^m \in \mathbb{R}^{|U| \times A}$ are the item-hyperedge and user-hyperedge dependency matrices. $\mathbf{E}_i^m$ is the original item feature matrix, $\mathbf{V}^m = \left[ \mathbf{v}_1^m, \ldots, \mathbf{v}_A^m \right] \in \mathbb{R}^{A \times d_m}$ is the hyperedge vector matrix, and $\mathbf{A}_u \in \mathbb{R}^{|{\cal U}| \times |\mathcal{I}|}$ is the user-related adjacency matrix extracted from $\mathbf{A}$.

During the construction of hyperedges, it is highly likely to make many discrete choices, such as deciding which items should link to which hyperedges. This can lead to meaningless connections and instability. Therefore, to mitigate such negative effects and improve the model’s generalization ability, Gumbel-Softmax is used for parameterization:
\begin{align}
\tilde{\mathbf{h}}_{i,*}^m = \text{SOFTMAX} \left( \frac{\log \bm{\delta} - \log (1 - \bm{\delta}) + \mathbf{h}_{i,*}^m}{\tau} \right),
\end{align}
where $\mathbf{h}_{i,*}^m \in \mathbb{R}^A$ is the $i$-th row vector of $\mathbf{H}_i^m$ representing the relationships between item $i$ and all hyperedges. $\bm{\delta} \in \mathbb{R}^A$ denotes a noise vector where each value $\bm{\delta}_j$ is drawn from a Uniform(0, 1) distribution, and $\tau$ signifies a temperature hyperparameter. Through the optimization steps mentioned above, we obtain the enhanced project hyperedge embedding $\tilde{\mathbf{h}}_i^m$. Similarly, applying the same operation to $\mathbf{H}_u^m$ yields the enhanced user hyperedge embedding $\tilde{\mathbf{H}}_u^m$. Then, by centering on hyperedge attributes, the embedding vectors are updated from layer \textit{h} to layer $\textit{h}+1$ by calculating the dot product of the relationship matrix and its transpose, then multiplying the result by the current embedding vectors. This process facilitates the transfer of information within the hypergraph conveying local information effectively to both users and items:
\begin{align}
\mathbf{E}_i^{m, h+1} = \text{DROP}(\tilde{\mathbf{H}}_i^m) \cdot \text{DROP}(\tilde{\mathbf{H}}_i^{m^\top}) \cdot \mathbf{E}_i^{m, h},
\end{align}
where $\mathbf{E}_i^{m,h}$ denotes the local embedding matrix of items in the \textit{h}-th hypergraph layer, and $\text{DROP}(\cdot)$ represents a dropout function. For the initial local embedding matrix when \textit{h} = 0, the collaborative embedding matrix $\mathbf{E}_{i, id}^{lge}$ of items is used. Additionally, the local user embedding matrix can be computed as follows:
\begin{align}
\mathbf{E}_u^{m,h+1} = \text{DROP}(\tilde{\mathbf{H}}_u^m) \cdot \text{DROP}(\tilde{\mathbf{H}}_i^{m^\top}) \cdot \mathbf{E}_i^{m,h}.
\end{align}

\begin{figure}
    \centering
    \includegraphics[width=1\linewidth]{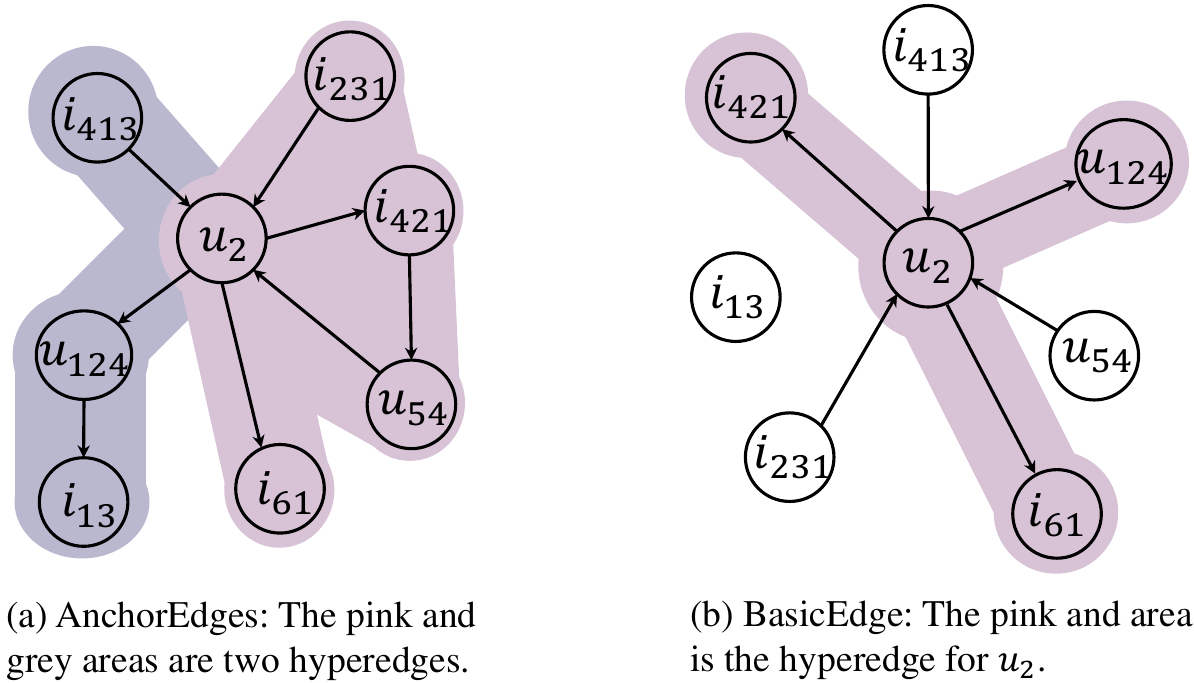}
    \caption{Visual Contrast: AnchorEdges and BasicEdges.}
    \label{fig:hypergraph}
\end{figure}

Ultimately, by merging the project embedding matrix with the user-project embedding as dependencies for the initial hypergraph, we explicitly achieve the integration of local embeddings from different modalities. This leads to the final local embedding matrix for the hypergraph segment $\mathbf{E}_{ghe}$:
\begin{align}
\mathbf{E}_{ghe} = \sum_{m \in \mathcal{M}} \mathbf{E}^{m,H}, \quad \mathbf{E}^{m,H} = \left[ \mathbf{E}_u^{m,H}, \mathbf{E}_i^{m,H} \right],
\end{align}
where $\mathbf{E}_u^{m,H} \in \mathbb{R}^{|{\cal U}| \times d}$ and $\mathbf{E}_i^{m,H} \in \mathbb{R}^{|\mathcal{I}| \times d}$ are the local embedding matrices of user $u$ and item $i$, respectively, obtained in the \textit{H}-th hypergraph layer under modality $m$.

\begin{algorithm}[!t]
\caption{The overall process of transformer layer construction.}
\label{alg:multihead_attention}
\textbf{Input}: User item interaction matrix $\mathbf{G}$, Embedding of user and items $\mathbf{E}$, number of multi-head layer $K$\\
\textbf{Parameter}: Learning weight matrices $\mathbf{W}^Q$, $\mathbf{W}^K$, $\mathbf{W}^V$\\
\textbf{Output}: The embedding matrices $\mathbf{resE}$ and Attention weight matrix $\mathbf{att}$
\begin{algorithmic}[1] %[1] enables line numbers
\STATE Initialize user item embedding: $\textbf{rows}$ and $\textbf{cols}$
\FOR{$k=0$ to $K$}
{
    \STATE $\mathbf{Q} = \text{LINEAR}(\textbf{rows}, \mathbf{W}^Q)$ 
    \STATE $\mathbf{K} = \text{LINEAR}(\textbf{cols}, \mathbf{W}^K)$
    \STATE $\mathbf{V} = \text{LINEAR}(\textbf{cols}, \mathbf{W}^V)$ \hfill \texttt{// Linear transformation and update iteration}
}
\ENDFOR
\STATE $\mathbf{pre\_att} = \text{SIMILARITY}(\mathbf{Q}, \mathbf{K}) = \frac{\mathbf{QK}^\top}{\sqrt{d_k}}$ \hfill \texttt{// Calculate the attention scores}
\STATE $\mathbf{pre\_att} = \text{SOFTMAX}\left(\frac{\mathbf{QK}^\top}{\sqrt{d_k}}\right)$ \hfill \texttt{// Apply softmax to attention scores}

\STATE $\mathbf{att} = \text{NORM}(\mathbf{pre\_att}, \textbf{cols})$  \hfill \texttt{// Convert attention scores to attention weights}
\STATE $\mathbf{att} = \text{NORM}(\mathbf{pre\_att}, \textbf{cols})$ \hfill \texttt{// compute final attention weight matrix}
\STATE $\mathbf{resE} = \text{ATTENTION}(\mathbf{att}, \mathbf{V}) = \text{SOFTMAX}\left(\frac{\mathbf{QK}^\top}{\sqrt{d_k}}\right) \mathbf{V}$ \hfill \texttt{// Compute the final embeddings}
\RETURN $\mathbf{att}, \mathbf{resE}$
\end{algorithmic}
\end{algorithm}

To further promote the integration of local information from different modalities, a self-supervised task is assigned to the cross-modal hypergraph fusion part. By treating the local embeddings of the same user as positive pairs and various users as negative pairs, the loss for the user hypergraph component is defined using InfoNCE, thus facilitating the alignment of distinct modalities:
\begin{align}
\mathcal{L}_{\text{HCL}}^u = \sum_{u \in \mathcal{U}} - \log \frac{\exp\left( s\left( \mathbf{E}_u^{v,H}, \mathbf{E}_u^{t,H} \right) / \tau \right)}{\sum_{{u'} \in \mathcal{U}} \exp\left( s\left( \mathbf{E}_{u'}^{v,H}, \mathbf{E}_{u'}^{t,H} \right) / \tau \right)},
\end{align}
where $s(\cdot)$ is the cosine similarity function, and $\tau$ is the temperature factor, typically set to 0.2. Note that only visual and textual modalities are considered here. Similarly, the item-side cross-modal contrastive loss $\mathcal{L}_{\text{HCL}}^i$ can be defined.

Incorporating self-supervised learning into the hypergraph construction process generates dynamic hyperedge structures, thereby improving the model's representational capacity. As illustrated in Fig. \ref{fig:hypergraph}, the gray section represents hyperedges formed through self-supervised learning which reveals the potential similarity relationships between data points, while the pink section corresponds to hyperedges constructed from inherent interaction patterns within the data. Unlike conventional methods that depend on manually labeled data to establish static hyperedges based on entity relationships, self-supervised learning empowers the model to utilize unlabeled data for representation learning. This is achieved through adaptive partial masking and prediction facilitating the dynamic generation of hyperedges. Consequently, this approach eliminates the need for manual labeling offering enhanced flexibility and adaptability.

\subsubsection{Transformer Layer Constructing}
Based on the calculation of the local structure of users under different modalities in the hypergraph, a transformer is introduced to further extract the global structure from user-item collaborative signals. However, introducing the transformer directly can cause unnecessary computations increasing the model’s time complexity. Additionally, experiments show that certain transformer components can lead to incorrect representations of embedding vectors reducing the model’s prediction accuracy. Therefore, the Transformer is modified to retain only its key component—the multi-head attention mechanism—to process user-item interaction signals. This assigns an attention weight to each user for different items allowing users to have varying sensitivities to various items. This approach captures the dynamic changes in user preferences as the number of interacted items increases, revealing the global behavior patterns of users and learning their global structure. The process of the multi-head attention mechanism can be represented as follows:
\begin{align}
\bm{\alpha}_{k,k'}^h = \frac{\exp \tilde{\bm{\alpha}}_{k,k'}^h}{\sum_{k'}\exp \tilde{\bm{\alpha}}_{k,k'}^h}  ;  \tilde{\bm{\alpha}}_{k,k'}^h = \frac{(\mathbf{W}_Q^h \cdot \bar{\mathbf{h}}_{k})^\top \cdot (\mathbf{W}_K^h \cdot \bar{\mathbf{h}}_{k'})}{\sqrt{d/H}},
\end{align}
Where $\bm{\alpha}_{k,k'}^h$ represents the attention weight between positions $k$ and $k'$ in the \textit{h}-th layer, while $\tilde{\bm{\alpha}}_{k,k'}^h$ denotes the unnormalized attention score between these positions. Note that $\mathbf{W}_Q^h$ and $\mathbf{W}_K^h$ are the query and key weight matrices for the \textit{h}-th head, respectively, and $\bar{\mathbf{h}}_k$ is the input vector at position $k$. Additionally, $d$ refers to the dimension of the input vectors and $H$ is the number of attention heads. Leveraging the calculated weights, they are mapped to the user-item collaborative signals through matrix dot multiplication, thereby assigning weights to the items interacted with by each user:
\begin{align}
\mathbf{E}_{GT} = {\bm{\alpha}}_{k,k'}^h \mathbf{E}_{c},
\end{align}
here, $\mathbf{E}_{GT}$ represents the global embedding, ${\bm{\alpha}}_{k,k'}^h$ denotes the attention weights, and $\mathbf{E}_{c}$ is the user-item collaborative embedding matrix. The above process is summarized in Algorithm \ref{alg:multihead_attention}.

The global structure and local structure learn from each other by combining the user's global behavior model with behaviors in specific domains allowing the recommendation system to dynamically adapt to changes in user interests. This enables more accurate and comprehensive predictions of user interest in new items improving recommendation precision. At the same time, to ensure a balance between global and local structures in user attention, we assign weights to both:
\begin{align}
\mathbf{E}_{str} = \alpha \cdot \text{NORM}(\mathbf{E}_{GT}) + \beta \cdot \text{NORM}(\mathbf{E}_{ghe}),
\end{align}
where $\text{NORM}(\cdot)$ denotes a normalization function to alleviate the value scale difference among embeddings, $\mathbf{E}_{GT}$ represents the global embedding, $\mathbf{E}_{ghe}$ signifies the local embedding and $\mathbf{E}_{str}$ conveys the refined structural embedding. $\alpha$ and $\beta$ are the proportion weights of the global embedding and local embedding, respectively.

\subsection{Fusion and Prediction}
To obtain the final representations $\mathbf{E}^*$ for users and items, their two types of embeddings are combined—the structurally refined embeddings $\mathbf{E}_{str}$ and the foundational multimodal collaborative signal embeddings $\mathbf{E}^{id}_{lge}$:
\begin{align}
\mathbf{E}^* = \mathbf{E}^{id}_{lge} + \mathbf{E}_{str}.
\end{align}

Then, the preference score $\hat{r}_{u,i}$ of user $u$ for item $i$ is calculated using the inner product, expressed as $\hat{r}_{u,i} = \mathbf{e}^{*}_{u} \cdot \mathbf{e}^{* \top}_{i}$. To optimize the model parameters, the Bayesian personalized ranking (BPR) loss function is employed:
\begin{align}
\mathcal{L}_{BPR} = -\sum_{(u,i^+,i^-) \in \mathcal{R}} \ln \sigma(\hat{r}_{u,i^+} + \hat{r}_{u,i^-}) + \lambda_1 \| \Theta \|_2^2,
\end{align}
here, $\mathcal{R} = \{(u, i^+, i^-)\ |\ (u, i^+) \in \mathcal{G},\ (u, i^-) \notin \mathcal{G}\}$ represents a set of triples used for training. $\sigma(\cdot)$ denotes the sigmoid function, while $\lambda_1$ and $\Theta$ refer to the regularization coefficient and the model parameters, respectively. Finally, the hypergraph contrastive loss, BPR loss, and multi-modal loss are unified into a single framework, as shown below:
\begin{align}
\mathcal{L} = \mathcal{L}_{BPR} + \lambda_2 \cdot (\mathcal{L}^{u}_{HCL} + \mathcal{L}^{i}_{HCL}) + \gamma \cdot (\mathcal{L}^{u}_{MCL} + \mathcal{L}^{i}_{MCL}),
\end{align}
here, $\lambda_2$ is the hyperparameter for the loss term weight, and $\gamma$ is the hyperparameter for the multi-modal loss weight. The Adam optimizer is employed to minimize the joint objective $\mathcal{L}$. A weight decay regularization term is applied to the model parameters $\Theta$.

\begin{table}[t]
\renewcommand{\arraystretch}{1.2}
\caption{Statistics of the three evaluation datasets.}
\resizebox{0.47\textwidth}{!}{%
\begin{tabular}{c|cccc}
\hline
\textbf{Dataset} & \textbf{\# User} & \textbf{\# Item} & \textbf{\# Interaction} & \textbf{Sparsity} \\ \hline
\textbf{Baby}     & 19,445 & 7,050  & 160,792 & 99.883\% \\
\textbf{Sports}   & 35,958 & 18,357 & 296,337 & 99.955\% \\
\textbf{Clothing} & 39,387 & 23,033 & 278,677 & 99.969\% \\ \hline
\end{tabular}%
}
\label{datasets}
\end{table}

Based on the above calculations, the resulting data matrix can be used to predict the score $\hat{r}_{u, i}$ for user $u$ selecting item $i$ through the prediction function, which is expressed by the following formula:
\begin{align}
\hat{r}_{u,i} = \text{PREDICTION}(\mathbf{R}, \mathbf{E}^{id}, \{\mathbf{E}^{m}_i\}_{m \in \mathcal{M}}),
\end{align}
where $\text{PREDICTION}(\cdot)$ is the prediction function. The matrix $\mathbf{E}^{id} = \left[ \mathbf{e}_{u_1}, \ldots, \mathbf{e}_{u_{|{\cal U}|}}, \mathbf{e}_{i_1}, \ldots, \mathbf{e}_{i_{|{\mathcal{I}}|}} \right] \in \mathbb{R}^{(|{\cal U}|+|{\mathcal{I}}|) \times d}$ represents the ID embedding matrix which stacks all the ID embeddings of users and items. The matrix $\mathbf{E}^{m}_{i} = \left[ \mathbf{e}^{m}_{i_1}, \ldots, \mathbf{e}^{m}_{i_{|{\mathcal{I}}|}} \right] \in \mathbb{R}^{|{\mathcal{I}}| \times d_{m}}$ denotes the item feature matrix under modality $m$. Moreover, $\mathcal{M}$ in this paper represents the collection of text and visual modalities.

\section{Experiments}
\label{section4}

In this section, an extensive series of experiments are conducted using three publicly available datasets to evaluate the effectiveness of the proposed SRGFormer model. The empirical findings address the following seven research questions:

\begin{itemize}
    \item RQ1: How does the performance of SRGFormer compare to other recommendation models? (Section \ref{section4B})
    \item RQ2: How does the inclusion of modal information affect the performance of SRGFormer? (Section \ref{section4C})
    \item RQ3: How do the main modules influence the performance of SRGFormer? (Section \ref{section4D})
    \item RQ4: How do key hyperparameters affect SRGFormer performance? (Section \ref{section4E})
    \item RQ5: How does the efficiency compare to other strong baselines? (Section \ref{section4F})
    \item RQ6: What is the resource consumption during the training of the entire model, and what potential limitations may arise? (Section \ref{section4G}) 
    \item RQ7: How does SRGFormer perform when tested on real-world datasets? (Section \ref{section4H})
\end{itemize}

\subsection{Experimental Settings}
\textbf{Datasets:} We use three public Amazon datasets: Baby, Sports, and Clothing, which include user-item interactions and item-specific data commonly used in multimedia recommendation research. 4096-dimensional raw visual features and 384-dimensional raw textual features extracted from prior studies are utilized. Detailed statistics are in Table \ref{datasets}.

\textbf{Evaluation Metrics:} We divide the historical interactions for each dataset into three parts: training, validation, and testing, using an 8:1:1 split. To assess the effectiveness of top-\textit{n} recommendation systems, two common evaluation metrics are implemented: Recall (R@\textit{n}) and Normalized Discounted Cumulative Gain (N@\textit{n}). The value of \textit{n} is varied between 10 and 20 and the average performance is calculated across all users in the test set.

% \usepackage{graphicx}
\begin{table}[t]
\caption{Parameter setting for three datasets.}
\tiny
\renewcommand{\arraystretch}{1}
\resizebox{0.45\textwidth}{!}{%
\begin{tabular}{c|cccc}
\hline
\textbf{Parameters} & \text{head} & $\alpha$ & $\beta$ & $\gamma$ \\ \hline
\textbf{Baby}       & 4             & 0.1            & 0.3            & 1e-06          \\
\textbf{Sports}     & 4             & 0.6            & 0.3            & 1e-06          \\
\textbf{Clothing}   & 4             & 0.2            & 0.4            & 1e-06          \\ \hline
\end{tabular}%
}
\label{hyperparameter}
\end{table}

\begin{table*}[!t]
\large
\renewcommand{\arraystretch}{1.1}
\caption{Performance comparison of different recommendation models. The optimal results are highlighted in \textbf{bold}, while the second-best results are \underline{underlined}.}
\resizebox{\textwidth}{!}{%
\begin{tabular}{lcccclcccclcccc}
\hline
\multicolumn{1}{c}{\multirow{2}{*}{\textbf{Model}}} & \multicolumn{4}{c}{\textbf{Baby}}                                     &  & \multicolumn{4}{c}{\textbf{Sports}}                                      &  & \multicolumn{4}{c}{\textbf{Clothing}}                                 \\ \cline{2-5} \cline{7-10} \cline{12-15} 
\multicolumn{1}{c}{}                                & \textbf{R@10}   & \textbf{R@20}   & \textbf{N@10}   & \textbf{N@20}   &  & \textbf{R@10}    & \textbf{R@20}   & \textbf{N@10}    & \textbf{N@20}    &  & \textbf{R@10}   & \textbf{R@20}   & \textbf{N@10}   & \textbf{N@20}   \\ \hline
\multicolumn{15}{l}{\textit{CF-based recommendation model}}                                                                                                                                                                                                                     \\ \hline
BPR                                                 & 0.0379          & 0.0607          & 0.0202          & 0.0261          &  & 0.0452           & 0.0690          & 0.0252           & 0.0314           &  & 0.0211          & 0.0315          & 0.0118          & 0.0144          \\ \hline
\multicolumn{15}{l}{\textit{Graph-based recommendation model}}                                                                                                                                                                                                                               \\ \hline
SGL                                                 & 0.0532          & 0.0820          & 0.0289          & 0.0363          &  & 0.0620           & 0.0945          & 0.0339           & 0.0423           &  & 0.0392          & 0.0586          & 0.0216          & 0.0266          \\
NCL                                                 & 0.0538          & 0.0836          & 0.0292          & 0.0369          &  & 0.0616           & 0.0940          & 0.0339           & 0.0421           &  & 0.0410          & 0.0607          & 0.0228          & 0.0275          \\
LightGCN                                            & 0.0464          & 0.0732          & 0.0251          & 0.0320          &  & 0.0553           & 0.0829          & 0.0307           & 0.0379           &  & 0.0331          & 0.0514          & 0.0181          & 0.0227          \\ \hline
\multicolumn{15}{l}{\textit{Hypergraph-based recommendation model}}                                                                                                                                                                                                                          \\ \hline
SHT                                                 & 0.0470          & 0.0748          & 0.0256          & 0.0329          &  & 0.0564           & 0.0838          & 0.0306           & 0.0384           &  & 0.0345          & 0.0541          & 0.0192          & 0.0243          \\
HCCF                                                & 0.0480          & 0.0756          & 0.0259          & 0.0332          &  & 0.0573           & 0.0857          & 0.0317           & 0.0394           &  & 0.0342          & 0.0533          & 0.0187          & 0.0235          \\ \hline
\multicolumn{15}{l}{\textit{Multimodal-based recommendation model}}                                                                                                                                                                                                                                \\ \hline
MMGCN                                               & 0.0398          & 0.0649          & 0.0211          & 0.0275          &  & 0.0382           & 0.0625          & 0.0200           & 0.0263           &  & 0.0229          & 0.0363          & 0.0118          & 0.0152          \\
VBPR                                                & 0.0424          & 0.0663          & 0.0223          & 0.0284          &  & 0.0556           & 0.0854          & 0.0301           & 0.0378           &  & 0.0281          & 0.0412          & 0.0158          & 0.0191          \\
GRCN                                                & 0.0531          & 0.0835          & 0.0291          & 0.0370          &  & 0.0600           & 0.0921          & 0.0324           & 0.0407           &  & 0.0431          & 0.0664          & 0.0230          & 0.0289          \\
LATTICE                                             & 0.0536          & 0.0858          & 0.0287          & 0.0370          &  & 0.0618           & 0.0950          & 0.0337           & 0.0423           &  & 0.0459          & 0.0702          & 0.0253          & 0.0306          \\
BM3                                                 & 0.0538          & 0.0857          & 0.0301          & 0.0378          &  & 0.0659           & 0.0979          & 0.0354           & 0.0437           &  & 0.0450          & 0.0669          & 0.0243          & 0.0295          \\
SLMRec                                              & 0.0540          & 0.0810          & 0.0296          & 0.0361          &  & 0.0676           & 0.1007          & 0.0374           & 0.0462           &  & 0.0452          & 0.0675          & 0.0247          & 0.0303          \\
MICRO                                               & 0.0570          & 0.0905          & 0.0310          & 0.0406          &  & 0.0675           & 0.1026          & 0.0365           & 0.0463           &  & 0.0496          & 0.0743          & 0.0264          & 0.0332          \\
LGMRec                                              & 0.0649         & 0.0989          & \underline{0.0353}          & \underline{0.0440}          &  & 0.0681           & 0.1044          & 0.0364           & 0.0458           &  & 0.0553          & 0.0827          & 0.0301          & 0.0374          \\
POWERec                                             & 0.0545          & 0.0823          & 0.0299          & 0.0370          &  & 0.0493           & 0.0765          & 0.0262           & 0.0332           &  & 0.0503          & 0.0753          & 0.0272          & 0.0336          \\ 
FREEDOM                                             & 0.0627          & 0.0989          & 0.0330         & 0.0424          &  & 0.0717           & 0.1089          & 0.0385           & 0.0481           &  & \underline{0.0629}          & \textbf{0.0941}          & \textbf{0.0341}          & \underline{0.0420}          \\
DRAGON                                             & \underline{0.0651}          & \underline{0.0991}          & 0.0328         & 0.0427          &  & \underline{0.0721}           & \underline{0.1092}          & \underline{0.0394}           & \underline{0.0496}           &  & \textbf{0.0631}          & \underline{0.0940}          & \underline{0.0338}          & \textbf{0.0424}          \\ \hline
\textbf{SRGFormer}                                  & \textbf{0.0681} & \textbf{0.1032} & \textbf{0.0369} & \textbf{0.0460} &  & \textbf{0.0758}  & \textbf{0.1146} & \textbf{0.0412}  & \textbf{0.0512}  &  & 0.0596 & 0.0884 & 0.0330 & 0.0403 \\
\textbf{$\triangle$ Improve.}                       & \textbf{4.61\%} & \textbf{4.14\%} & \textbf{4.53\%} & \textbf{4.54\%} &  & \textbf{5.14\%} & \textbf{4.95\%} & \textbf{4.57\%} & \textbf{3.23\%} &  & -5.54\% & -6.05\% & -3.22\% & -4.95\% \\ \hline
\end{tabular}
}
\label{performance}
\end{table*}

\textbf{Baseline Models:} To demonstrate the effectiveness of our proposed model, we perform a comparative analysis against several well-established recommendation models. These baseline models are categorized into four distinct groups: (1) collaborative filtering-based (CF-based) model, (2) graph-based recommendation model, (3) hypergraph-based recommendation model, and (4) multimodal-based recommendation model.

1) CF-based recommendation model:

\begin{itemize}
    \item BPR \cite{27} (UAI'09): This method optimizes the ranking of items by modeling the relative preferences of users, aiming to ensure that preferred items are ranked higher than non-preferred ones.
\end{itemize}

2) Graph-based recommendation model:
\begin{itemize}
    \item NCL \cite{40} (NeurIPS'18): This method boosts model generalization and robustness by training multiple classifier heads simultaneously, using consensus among them and sharing intermediate-level representations.
    \item LightGCN \cite{11} (SIGIR'20): This method focuses on the essential components of graph convolutional networks by eliminating unnecessary operations.
    \item SGL \cite{39} (SIGIR'21): This method uses self-supervised tasks to enhance the model's generalization ability by incorporating relational information from the graph.
\end{itemize}

3) Hypergraph-based recommendation model:
\begin{itemize}
    \item SHT \cite{41} (KDD'22): This method improves the robustness and effectiveness of graph-based collaborative filtering by combining hypergraph neural networks with Transformer-like mechanisms to tackle data sparsity.
    \item HCCF \cite{42} (SIGIR'22): This approach enhances the model's ability to more effectively grasp user preferences and item attributes.
\end{itemize}

4) Multimodal-based recommendation model:
\begin{itemize}
    \item VBPR \cite{24} (AAAI'16): This method leverages user-item interaction data and visual content to enhance the quality of recommendations.
    \item MMGCN \cite{28} (ACM MM'19): This method adeptly captures the intricate interactions among users and items across various modalities, leading to more personalized.
    \item GRCN \cite{43} (ACM MM'20): This method adaptively refines interaction graphs, enhancing multimedia recommendations by pruning false-positive edges.
    \item LATTICE \cite{38} (ACM MM'21): This approach employs a modal-aware structure learning layer to construct item graphs.
    \item SLMRec \cite{45} (IEEE TMM'22): This method integrates self-supervised learning into a multimedia recommendation by using data augmentation and contrastive learning.
    \item MICRO \cite{36} (CIKM'22): This method employs modal-preserving decoders to maintain each data type's unique properties. 
    \item BM3 \cite{44} (WWW'23): This method uses dropout for contrastive views and optimizes three objectives to align and reconstruct user-item interactions across modalities. 
    \item FREEDOM \cite{37} (ACM MM'23): This model boosts accuracy and efficiency by freezing the item-item graph, denoising the user-item graph, and integrating both for robust multimodal representation.
    \item DRAGON \cite{50} (ECAI'23): This model enhances recommendations by learning dual user-item representations through graph integration and attentive fusion of multimodal features.
    \item LGMRec \cite{46} (AAAI'24): This method uses graph learning to model local interactions and hypergraph learning to capture global dependencies.
    \item POWERec \cite{47} (Inform. Fusion'24): This method uses a single user embedding with modality prompts to capture user interests across different modalities. 
\end{itemize}

\textbf{Parameter Settings:} Our model is implemented in Pytorch and fine-tuned with essential parameters. For graph-based methods, the number of Collaborative Graph Propagation layers $s$ is set to 2. Model parameters are initialized using the Xavier method, and optimal hyperparameters are identified via grid search on the validation set. Specifically, the weights for local and global structure embeddings ($\alpha$ and $\beta$) are adjusted within $\{0.1, 0.2, \ldots, 1\}$. The multi-head attention mechanism layers ($\text{head}$) take values within $\{2, 4, \ldots, 12\}$, and the weights for multimodal self-supervised tasks ($\gamma$) are adjusted within $\{1 \times 10^{-8}, 1 \times 10^{-7}, \ldots, 0.1\}$. Training stops early if R@20 on the validation set does not improve for 20 consecutive epochs. The specific parameter values are provided in Table \ref{hyperparameter}.

\subsection{Performance Comparison (RQ1)}\label{section4B}
Table \ref{performance} summarizes the performance of various recommendation methods across three datasets, with key observations as follows: Firstly, SRGFormer significantly outperforms general and state-of-the-art multimodal techniques averaging improvements of 4.46\% on the Baby dataset, 4.47\% on the Sports dataset. Secondly, hypergraph models like HCCF and SHT that learn user structures do not always surpass general models, indicating they fail to capture global user structural patterns. Lastly, the poor performance of MMGCN and VBPR suggests two issues: they incorporate multimodal signals without further learning, and they study high-order user-item interactions without deeply learning user structural features, leading to uniform sensitivity to different items.

\begin{itemize}
    \item SRGFormer addresses these issues by using a transformer-based attention mechanism and a multimodal-based hypergraph to learn refined user structures jointly and by introducing self-supervised tasks for deep multimodal learning. This enhances interaction among modalities and improves recommendation accuracy. Compared to alternative attention-based models, SRGFormer's significant improvements demonstrate the potential of multimodal interactions in modeling modality-aware dependencies.

    \item SRGFormer performs sub-optimally on the clothing dataset highlighting its inherent limitations. The best-performing baseline model on this dataset, FREEDOM, mitigates the impact of large data volumes on model efficiency using "freezing" and "denoising" techniques. In contrast to FREEDOM, SRGFormer preserves the inherent connectivity of the graph structure. It instead learns the deep structure of user-item interactions by dynamically adapting to changes in the data enabling it to capture potential semantics at any moment. Additionally, the integration of self-supervised tasks and attention mechanisms offers a significant advantage in managing highly sparse interaction data. However, as the number of edges and nodes increases, the complexity and memory consumption become more significant leading to decreased efficiency. To address this, we will continue to refine and optimize these mechanisms ensuring better scalability for larger datasets.
\end{itemize}

\begin{figure*}[!t]
    \centering
    \includegraphics[width=1\linewidth, height=0.25\linewidth]{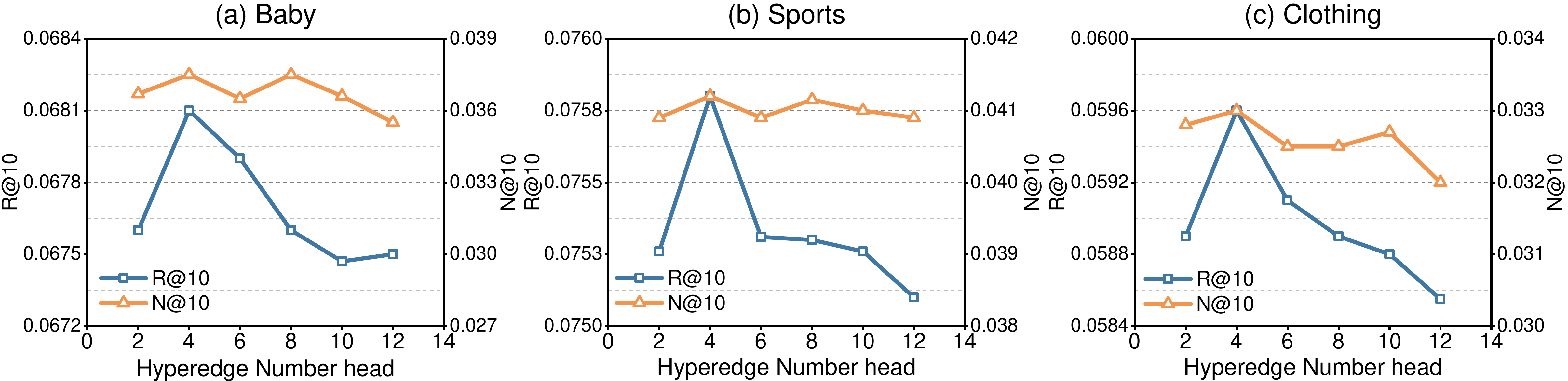}
    \caption{The performance of hyperparameter head on the Baby, Sports, and Clothing datasets in terms of Recall@10 and NDCG@10.}
    \label{fig:head}
\end{figure*}

\subsection{Effects of Modality (RQ2)}\label{section4C}
To explore how different modalities influence recommendation outcomes, an ablation study is conducted on the SRGFormer model. This study compares the full SRGFormer model with two variants: $\text{SRGFormer}_{w/t}$ which excludes text data, and $\text{SRGFormer}_{w/v}$ which excludes visual information. The goal is to isolate the impact of each modality on performance.

Table \ref{modility} shows SRGFormer's consistent superiority over its variants, particularly in the Sports dataset, where it achieves a performance boost of 11.31\% in Recall@10 and 12.88\% in NDCG@10 compared to $\text{SRGFormer}_{w/t}$. This highlights the benefits of a multimodal approach in accurately predicting user preferences. In the Clothing dataset, $\text{SRGFormer}_{w/v}$ shows a significant performance decline, underscoring the importance of visual information in sectors like fashion.

\begin{table}[!t]
\caption{Ablation study on the different modalities of SRGFormer.}
\setlength{\tabcolsep}{2.2pt}
\renewcommand{\arraystretch}{1.48}
\resizebox{0.48\textwidth}{!}{%
\begin{tabular}{cl|cc|cc|cc}
\hline
\multicolumn{2}{c|}{\textbf{Datasets}} & \multicolumn{2}{c|}{\textbf{Baby}} & \multicolumn{2}{c|}{\textbf{Sports}} & \multicolumn{2}{c}{\textbf{Clothing}} \\ \hline
\multicolumn{2}{c|}{\textbf{Metric}} & \textbf{R@10}   & \textbf{N@10}   & \textbf{R@10}   & \textbf{N@10}   & \textbf{R@10}  & \textbf{N@10}   \\ \hline
\multicolumn{2}{c|}{$\text{SRGFormer}_{w/v}$}            & 0.0641          & 0.0347          & 0.0745          & 0.0406          & 0.0571         & 0.0241          \\
\multicolumn{2}{c|}{$\text{SRGFormer}_{w/t}$}            & 0.0632          & 0.0341          & 0.0738          & 0.0399          & 0.0584         & 0.0301          \\ \hline
\multicolumn{2}{c|}{SRGFormer}            & \textbf{0.0681} & \textbf{0.0369} & \textbf{0.0758} & \textbf{0.0412} & \textbf{0.0596} & \textbf{0.0330} \\ \hline
\end{tabular}%
}
\label{modility}
\end{table}

\begin{table}[!t]
\caption{Performance comparison under different ablation datasets.}
\setlength{\tabcolsep}{2.2pt}
\renewcommand{\arraystretch}{1.48}
\resizebox{0.48\textwidth}{!}{%
\begin{tabular}{cl|cc|cc|cc}
\hline
\multicolumn{2}{c|}{\textbf{Datasets}} & \multicolumn{2}{c|}{\textbf{Baby}} & \multicolumn{2}{c|}{\textbf{Sports}} & \multicolumn{2}{c}{\textbf{Clothing}} \\ \hline
\multicolumn{2}{c|}{\textbf{Metric}} & \textbf{R@10}   & \textbf{N@10}   & \textbf{R@10}   & \textbf{N@10}   & \textbf{R@10}  & \textbf{N@10}   \\ \hline
\multicolumn{8}{l}{\textit{Recent-behavior Masked Dataset (RBM-D)}}  \\ \hline
\multicolumn{2}{c|}{$\text{LightGCN}$}            & 0.0453          & 0.0243          & 0.0539          & 0.0289          & 0.0323         & 0.0167          \\
\multicolumn{2}{c|}{$\text{LGMRec}$}            & 0.0641          & 0.0346          & 0.0672          & 0.0357          & 0.0547         & 0.0298          \\ 
\multicolumn{2}{c|}{$\text{SRGFormer}$}            & 0.0677          & 0.0356          & 0.0747          & 0.0405          & 0.0584         & 0.0322         \\ 
\multicolumn{2}{c|}{$\text{SRGFormer}_{w/h}$}            & 0.0651          & 0.0339          & 0.0736          & 0.0391          & 0.0562         & 0.0308   \\ \hline
\multicolumn{8}{l}{\textit{Long-history Masked Dataset (LHM-D)}}  \\ \hline
\multicolumn{2}{c|}{$\text{LightGCN}$}            & 0.0441          & 0.0231          & 0.0534          & 0.0285          & 0.0311         & 0.0152          \\
\multicolumn{2}{c|}{$\text{LGMRec}$}            & 0.0643          & 0.0341          & 0.0673          & 0.0353          & 0.0545         & 0.0296          \\
\multicolumn{2}{c|}{$\text{SRGFormer}$}            & 0.0674          & 0.0356          & 0.0749          & 0.0407          & 0.0589         & 0.0324         \\ \multicolumn{2}{c|}{$\text{SRGFormer}_{w/h}$}            & 0.0659          & 0.0343          & 0.0739          & 0.0398          & 0.0572         & 0.0314   \\ \hline
\multicolumn{8}{l}{\textit{Full-information Dataset (FID)}}  \\ \hline
\multicolumn{2}{c|}{$\text{LightGCN}$}            & 0.0464          & 0.0251          & 0.0553          & 0.0307          & 0.0331         & 0.0181          \\ 
\multicolumn{2}{c|}{$\text{LGMRec}$}            & 0.0649          & 0.0353          & 0.0681          & 0.0364          & 0.0553         & 0.0301          \\
\multicolumn{2}{c|}{$\text{SRGFormer}$}            & 0.0681          & 0.0369          & 0.0758          & 0.0412          & 0.0596         & 0.0330     \\ \hline
\end{tabular}%
}
\label{local-global}
\end{table}

Overall, the results confirm SRGFormer's effectiveness in utilizing multimodal information and the critical importance of each modality for accurate recommendations. By leveraging text and visual inputs, SRGFormer sets a new benchmark in recommendation systems, demonstrating how multimodal data integration enhances decision-making across various consumer environments. This analysis supports multimodal data as essential for next-generation recommendation systems, ensuring they are adaptive and contextually relevant.

\subsection{Key Components (RQ3)}\label{section4D}
To assess the significance of specific components in the SRGFormer framework, we analyze two variants: (1) $\text{SRGFormer}_{w/GT}$ which excludes the multi-head attention layer, and (2) $\text{SRGFormer}_{w/MCL}$ which omits the self-supervised task for multimodal interactions. These modifications aim to isolate and evaluate the contributions of these key elements to the model's performance.

The performance impacts of these variants are documented in Table \ref{key components}. The variant without the multi-head attention layer, $\text{SRGFormer}_{w/GT}$, shows a substantial decrease in effectiveness across all three datasets, highlighting the crucial role of the multi-head attention layer in processing user information sensitively and dynamically. Without it, the model struggles to discern and adapt to complex user preferences.

In contrast, the $\text{SRGFormer}_{w/MCL}$ variant which removes the self-supervised task for enhancing multimodal interactions, shows a less severe performance decline. Although this variant can function without the self-supervised component, it is less efficient in capturing and integrating inter-modal information leading to less effective data representation and failing to fully leverage the knowledge within diverse modal information about users or items.

To evaluate local-global information learning, we process the dataset in three ways: (1) \textit{RBM-D}, where the most recent k actions are removed to test local reliance; (2) \textit{LHM-D}, where only the most recent L actions are kept to test global reliance; and (3) \textit{FID}, the original dataset. Performance differences between LightGCN (without local-global learning) and LGMRec / SRGFormer (with local-global learning) across these dataset settings are analyzed to demonstrate the effectiveness of local-global learning.

From Table \ref{local-global}, we observe: (1) Compared to the two models based on local-global information learning, the traditional LightGCN performs worse on the processed dataset. Simultaneously, compared to the \textit{RBM-D} model designed for short-term behavior datasets, LightGCN exhibits smaller fluctuations on the long-term behavior dataset \textit{LHM-D}, indicating stronger robustness when handling global features; (2) The models LGMRec and SRGFormer, which learn from both local and global information, exhibit only minor performance fluctuations across different datasets, validating the importance of the local-global information learning module. (3) To analyze the hypergraph mechanism in SRGFormer, experiments on $\text{SRGFormer}_{w/h}$ under modified datasets show larger performance drops when local information is masked, revealing its dependence on local structural cues. With our constrained hyperedge design, the model achieves stronger semantic consistency within local groups. Unlike LightGCN, the hypergraph aggregates contextual signals from entire groups in one step, mitigating structural distortion caused by multi-hop diffusion.

\subsection{Hyperparameter Analysis (RQ4)}\label{section4E}
\begin{figure*}[!t]
    \centering
    \includegraphics[width=1\linewidth]{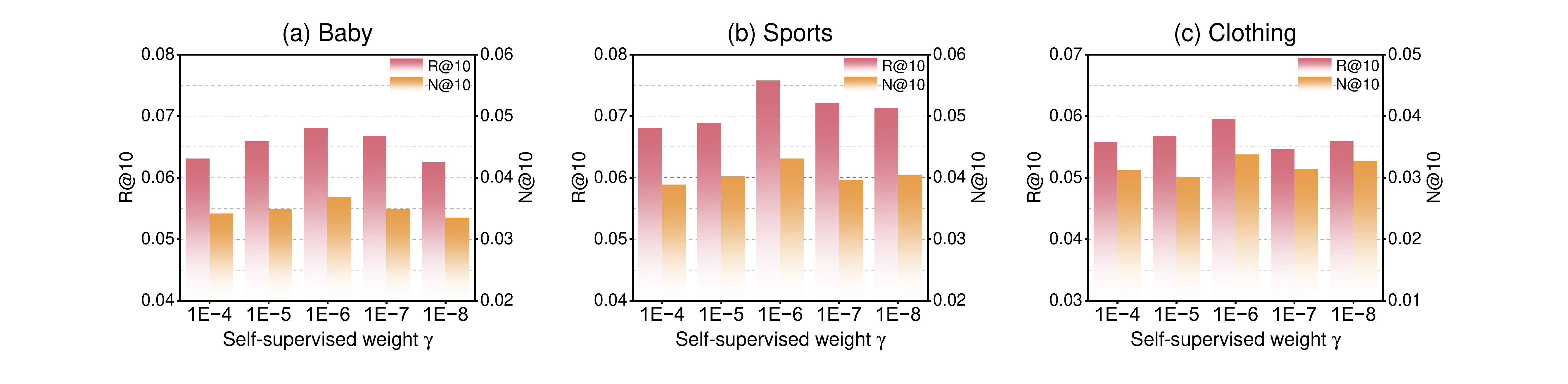}
    \caption{The performance of hyperparameter $\gamma$ on the Baby, Sports, and Clothing datasets in terms of Recall@10 and NDCG@10.}
    \label{fig:gama}
\end{figure*}

To verify the effectiveness of two key hyperparameters ($\text{head}$ and $\gamma$), we conduct experiments on three datasets using the control variable method. The experimental results, as shown in Fig. \ref{fig:head} and Fig. \ref{fig:gama}, depict the performance variations of Recall@10 and NDCG@10 on the three datasets.

The results indicate that for the number of heads in the multi-head attention layers, too many attention layers fail to explore user preferences deeply and performance decreases due to increased time consumption after learning saturates. Conversely, too few attention layers are insufficient for deeply learning user preferences. Observations on the weight of the self-supervised task $\gamma$ are similar, showing a trend of first increasing and then decreasing. The figure illustrates that an appropriate $\gamma$ can improve accuracy through sufficient self-supervised tasks, but an excessively high $\gamma$ may hurt performance.

\begin{table}[!t]
\caption{Ablation study on key components of SRGFormer.}
\setlength{\tabcolsep}{2.2pt}
\renewcommand{\arraystretch}{1.58}
\resizebox{0.48\textwidth}{!}{%
\begin{tabular}{cl|cc|cc|cc}
\hline
\multicolumn{2}{c|}{\textbf{Datasets}} & \multicolumn{2}{c|}{\textbf{Baby}} & \multicolumn{2}{c|}{\textbf{Sports}} & \multicolumn{2}{c}{\textbf{Clothing}} \\ \hline
\multicolumn{2}{c|}{\textbf{Metric}} & \textbf{R@10}   & \textbf{N@10}   & \textbf{R@10}   & \textbf{N@10}   & \textbf{R@10}  & \textbf{N@10}   \\ \hline
\multicolumn{2}{c|}{$\text{SRGFormer}_{w/GT}$}            & 0.0644          & 0.0339          & 0.0720          & 0.0390          & 0.0555         & 0.0217          \\
\multicolumn{2}{c|}{$\text{SRGFormer}_{w/MCL}$}            & 0.0672          & 0.0353          & 0.0741          & 0.0406          & 0.0574         & 0.0289          \\ \hline
\multicolumn{2}{c|}{SRGFormer}            & \textbf{0.0681} & \textbf{0.0369} & \textbf{0.0758} & \textbf{0.0412} & \textbf{0.0596} & \textbf{0.0330} \\ \hline
\end{tabular}%
}
\label{key components}
\end{table}

\begin{figure*}[!t]
    \centering
    \includegraphics[width=0.95\linewidth]{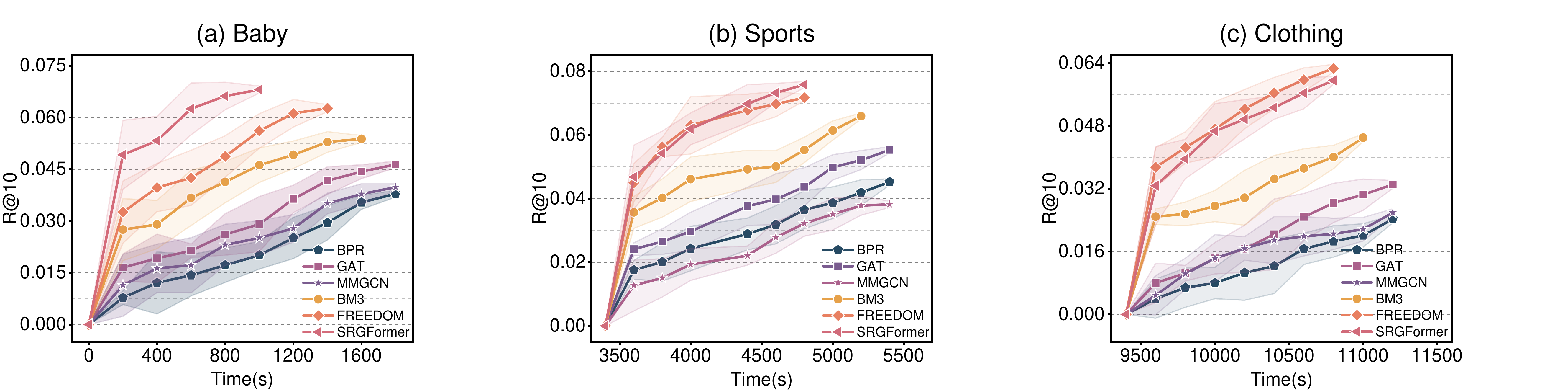}
    \caption{The efficiency performance of SRGFormer is compared to various baseline models on the Baby, Sports, and Clothing datasets in terms of Recall@10.}
    \label{fig:time}
\end{figure*}

\subsection{Efficiency Benchmarking (RQ5)}\label{section4F}
To evaluate SRGFormer's efficiency, we compare it with several state-of-the-art models, recording their respective performance metrics at fixed time intervals to ensure a comprehensive assessment. Specifically, we compare three classical models to provide a thorough evaluation of the efficiency of our model: the CF-based recommendation model, the Graph-based recommendation model, and the Multimodal-based recommendation model. The specific time efficiency is shown in the Fig. \ref{fig:time}.

For smaller datasets, such as Baby and Sports, SRGFormer balances efficiency and accuracy. Compared to existing models, the multi-head attention mechanism aggregates information by leveraging parallel processing and information sharing, effectively reducing computational load. Compared to the traditional GAT, SRGFormer selectively integrates the Transformer architecture and removes modules with lower relevance to the recommendation environment (e.g., FFN), thereby reducing unnecessary computations and improving model efficiency and performance. Furthermore, the use of self-supervised hypergraphs reduces reliance on manual annotations by dynamically learning the data structure and optimizing the filtering of useful information. Thanks to these optimizations, SRGFormer performs well early in training and continues progressing steadily thereafter.

When handling larger datasets, such as the Clothing dataset, SRGFormer's efficiency and performance decline due to the challenges of handling large volumes of graph and node data resulting in an exponential increase in learning complexity and computational cost. In contrast, FREEDOM’s effective “freezing” and “denoising” mechanisms enable it to surpass our model on large-scale datasets. This highlights a limitation of our approach, which will be addressed in future research.

\subsection{Cost Analysis and Key Challenges (RQ6)}\label{section4G}
The training is performed throughout the experiment on the 24GB memory of the NVIDIA GeForce RTX 3090. The computational load of our SRGFormer model arises from two main components: (1) the Transformer Layer Constructing module and (2) the Hypergraph Constructing module.

On the one hand, the hypergraph construction module processes and integrates multimodal information to build the hypergraph structure. The computation of hyperedge dependencies involves dot product operations and a self-supervised process. Suppose the number of hyperedges is denoted as $E$ and the input matrix has dimensions $[n, m]$. In that case, the time complexity is $\mathcal{O}(E*n*m*d*L+n*m*d)$ where $L$ is the number of hypergraph layers, $d$ is the embedding dimension and $n$ and $m$ correspond to the discretized user-user (item-item) graph in this model. On the other hand, the multi-head attention mechanism aggregates information. It assigns weights with its computational cost primarily stemming from dot product and softmax operations particularly the transformations between the $Q$, $K$, and $V$ matrices. Assuming the dimensions of these matrices are $[n, m]$, the time complexity is $\mathcal{O}(h*n^2*m)$ where $h$ represents the number of attention heads, $n$ and $m$ correspond to the number of user and item nodes derived from the normalized user-item matrix.

In summary, the model's computational efficiency and overhead are primarily influenced by the number of edges in the graph followed by the number of attention heads and nodes. As the dataset grows, the load increases due to hyperedge constructions and multiple head computations. Consequently, training and inference on smaller datasets are less resource-demanding whereas larger datasets slow convergence affecting efficiency and performance. Additionally, since self-supervised learning depends on extracting high-quality features from the data, insufficient data may cause the model to overfit the training data thereby decreasing its generalization capability.

\subsection{Case Study (RQ7)}\label{section4H}
Case studies were conducted on representative users from the Baby dataset to investigate SRGFormer's interpretability. We examine user reviews and item features not used in model training and check the sensitivity scores $\overline{\alpha}$ assigned by users. 

\begin{figure}[t]
    \centering
    \includegraphics[width=1\linewidth]{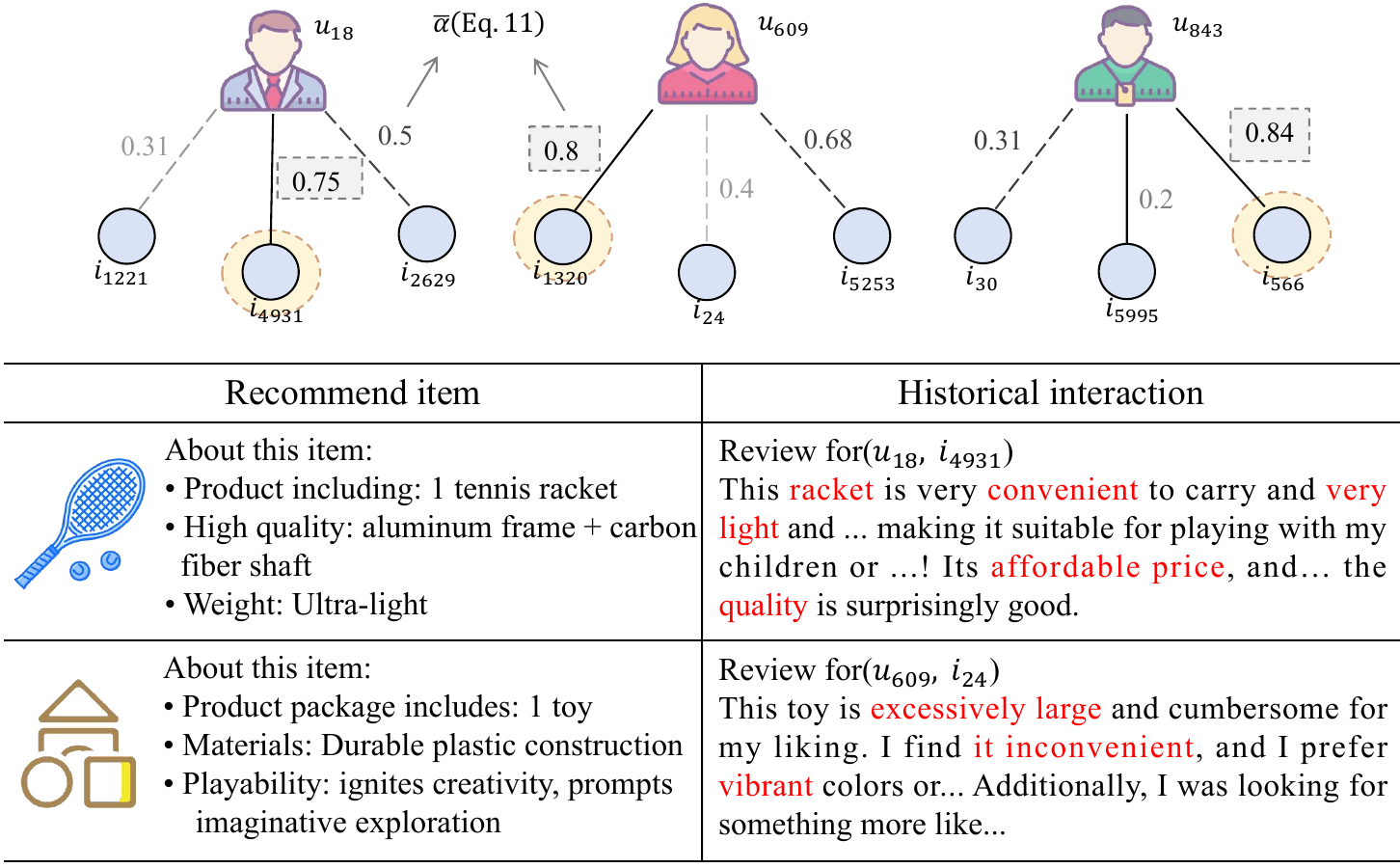}
    \caption{Case study on the principles of knowledge collaboration aims at extracting information that users truly care about from noisy interactions.}
    \label{case-study}
\end{figure}

As shown in Fig. \ref{case-study}, items with the highest sensitivity scores for users $u_{18}$, $u_{609}$, and $u_{843}$ match items with positive feedback while items with unsatisfactory reviews receive lower scores. These results indicate that SRGFormer effectively highlights user sensitivity to different items and reallocates sensitivity based on actual preferences. Examining $u_{18}$'s scores, the highest sensitivity score $(u_{18}, i_{4931})$ matches the user's highest rating aligning with the item's feature attributes and illustrating the importance of user sensitivity. For $u_{609}$, the lowest sensitivity score $(u_{609}, i_{24})$ matches the user's lowest rating indicating a mismatch with the item's attributes. This analysis shows that lower sensitivity reflects a lack of interest, validating the practicality and robustness of sensitivity in different scenarios.

\section{Conclusion}
\label{section5}
This paper refines user structures to enhance collaborative filtering by assigning different item weights through a transformer-based multi-head attention mechanism. The proposed SRGFormer model integrates users' global and local structures and explores multimodal interactions. This approach facilitates learning associations between different modalities via self-supervised tasks. Empirical results show that transformer-based multi-head attention and effective self-supervised task formulation significantly improve user preference learning and recommendation performance. In future work, we intend to investigate more attention variants to capture intricate information within multimodal features, aiming to improve the model's recommendation performance.

% References
\bibliographystyle{Tcyber/Bibliography/IEEEtran}
\bibliography{Tcyber/Bibliography/Bibfile}

\vspace{-25pt}

\begin{IEEEbiography}[{\includegraphics[width=1in, height=1.25in, clip, keepaspectratio]{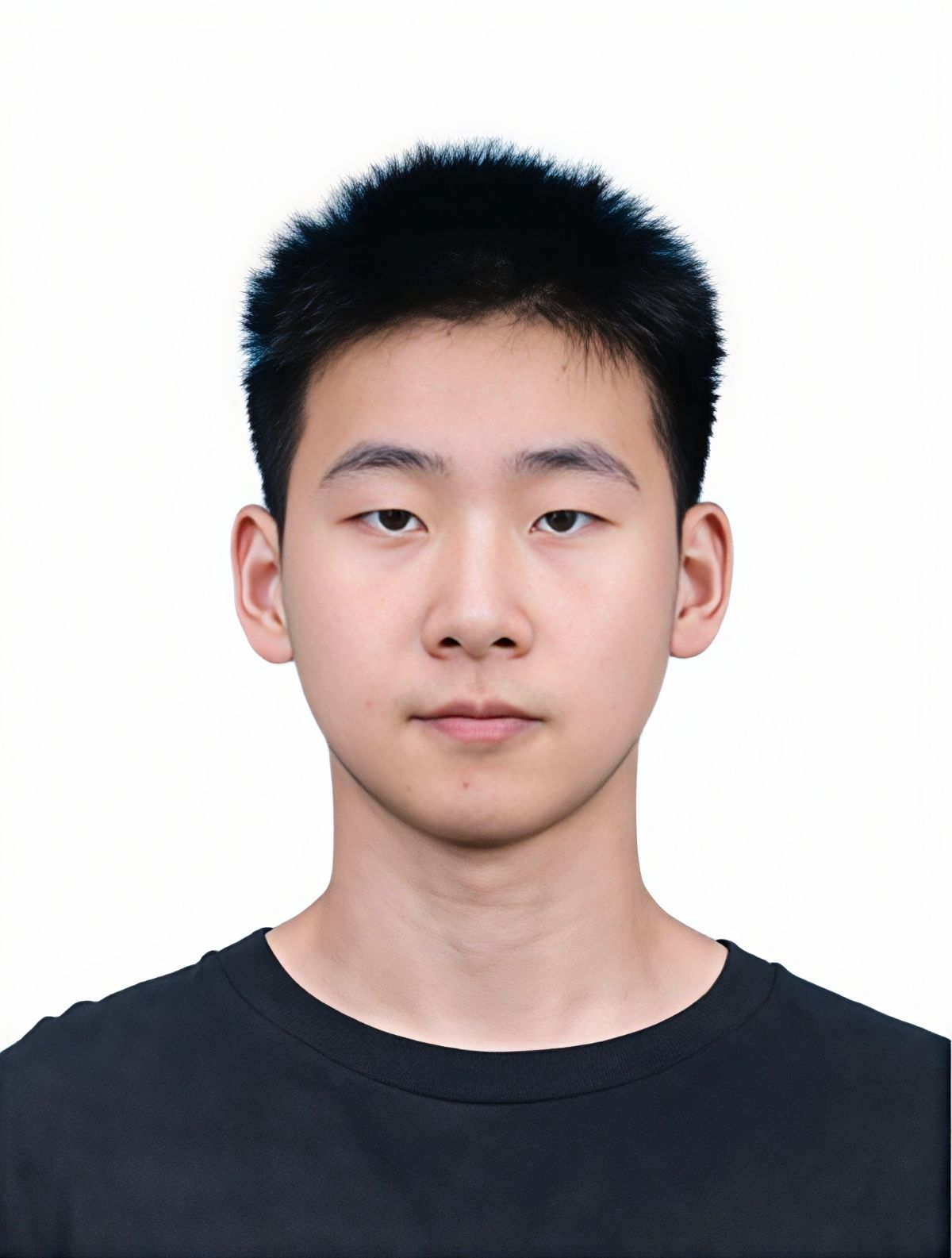}}] {Ke Shi} is currently pursuing the B.E. degree in Computer Science and Technology at Hubei University, Wuhan, China. His main research direction is recommendation systems.
\end{IEEEbiography}

\vspace{-25pt}

\begin{IEEEbiography}[{\includegraphics[width=1in, height=1.25in, clip, keepaspectratio]{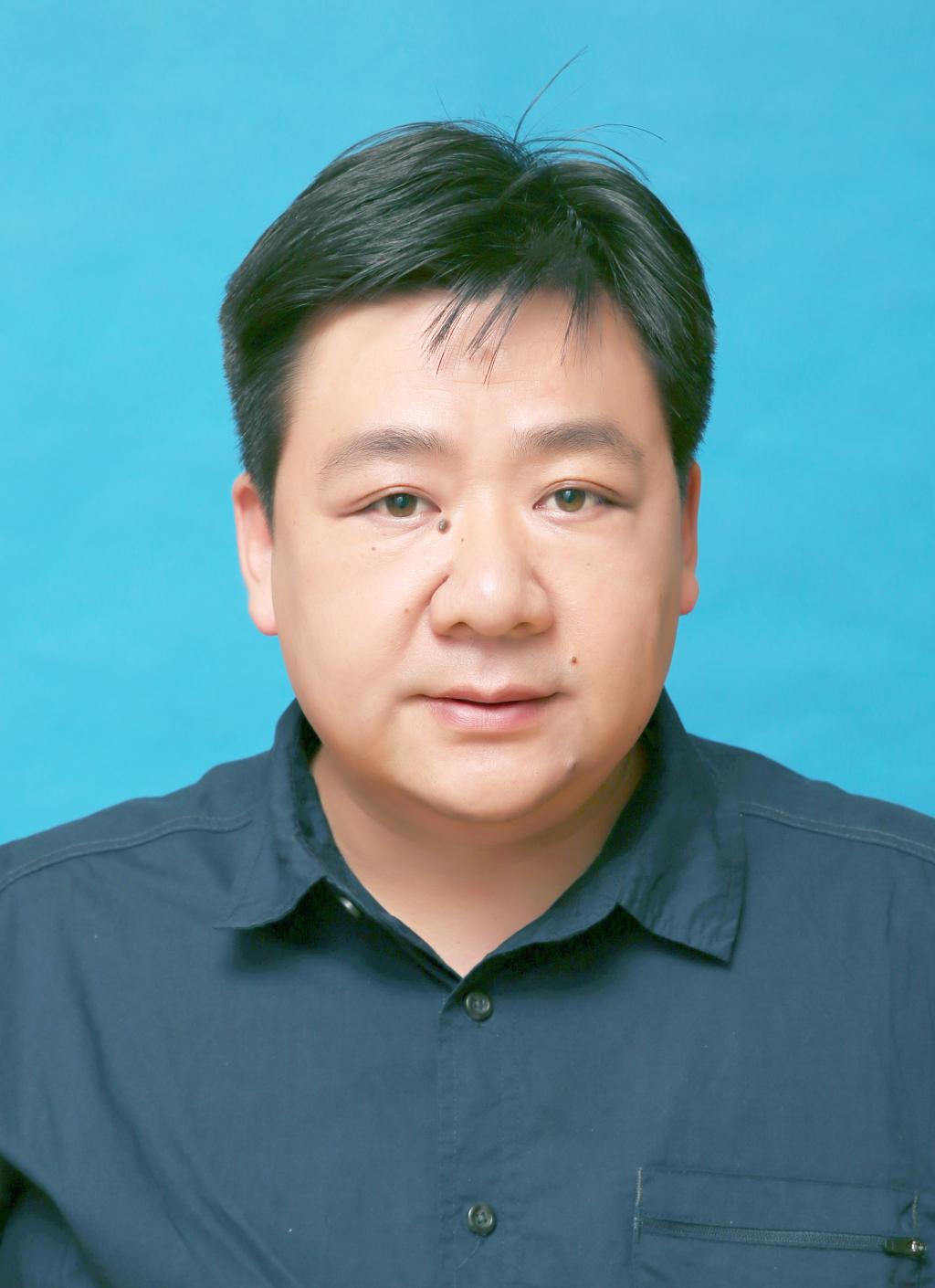}}]{Yan Zhang} received the Ph.D. degree from Beihang University. He is currently a professor in the School of Computer Science at Hubei University. His research interests include cyberspace security and software engineering, among which he has in-depth theoretical research and practical experience in source code security detection and verification.
\end{IEEEbiography}

\vspace{-25pt}

\begin{IEEEbiography}[{\includegraphics[width=1in, height=1.25in, clip, keepaspectratio]{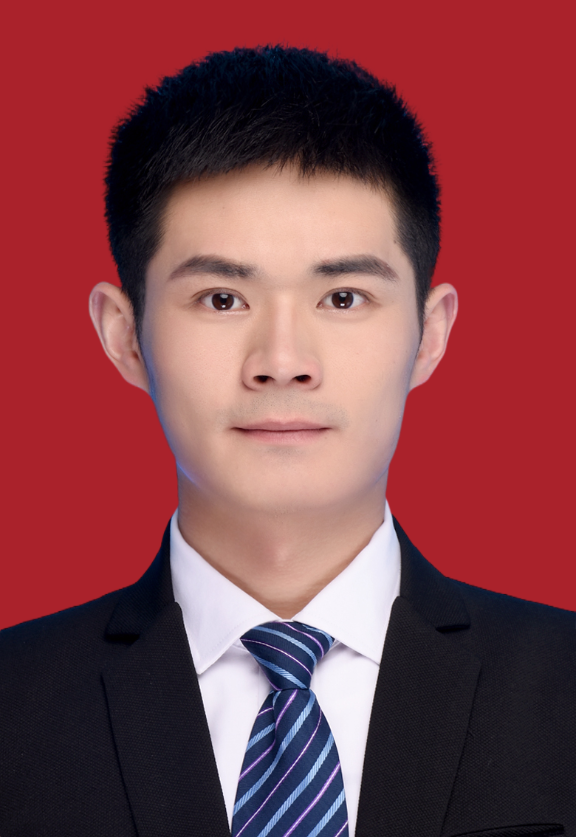}}]{Miao Zhang} received the Ph.D. degree from the National Engineering Research Center for E-Learning at Central China Normal University in 2023. He received the M.S. degree in computer technology from Central China Normal University, China in 2018. Since October 2025, he has been an Associate Professor at the School of Computer Science, Hubei University. His research interests include Q\&A systems and knowledge graphs.
\end{IEEEbiography}

\vspace{-25pt}

\begin{IEEEbiography}[{\includegraphics[width=1in, height=1.25in, clip, keepaspectratio]{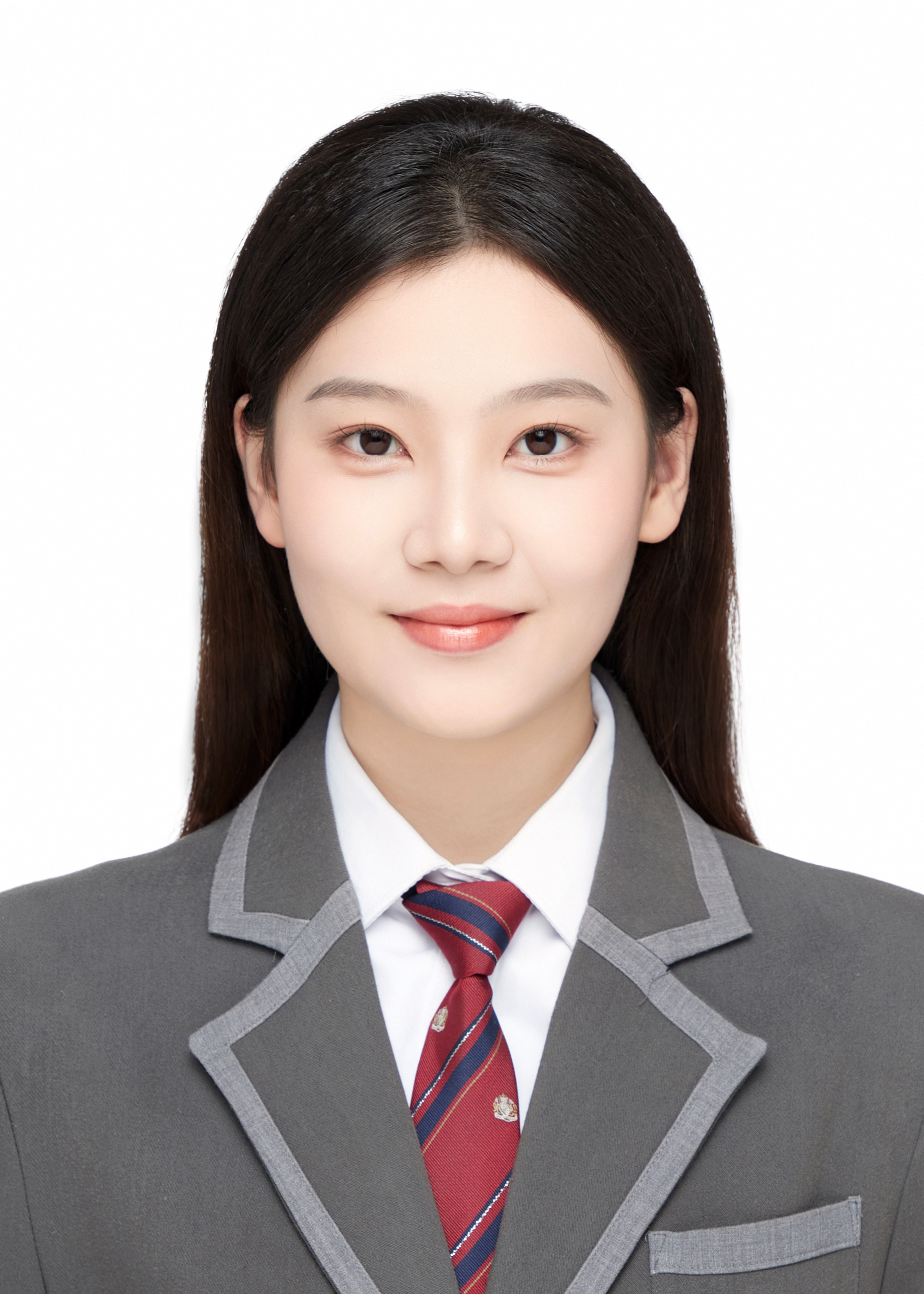}}] {Lifan Chen} received the B.E. degree in cyberspace security from Shandong University of Political Science and Law, Jinan, China, in 2024. She is currently pursuing the M.S. degree at Hubei University. Her research interests include deep learning and educational knowledge tracing.
\end{IEEEbiography}

\vspace{-25pt}

\begin{IEEEbiography}[{\includegraphics[width=1in, height=1.25in, clip, keepaspectratio]{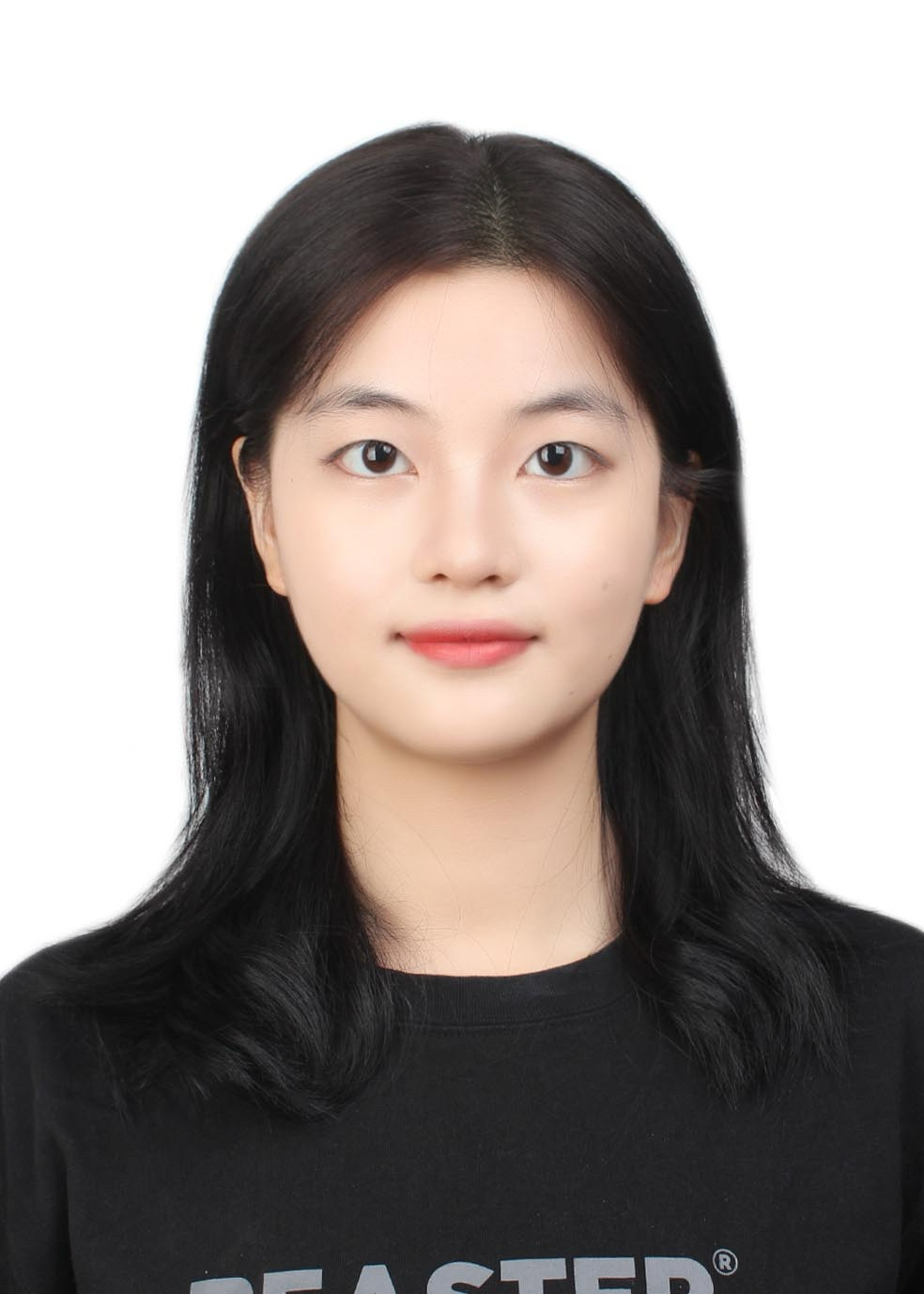}}] {Jiali Yi} received a B.E. degree in Wenhua College, Wuhan, China, in 2024. She is currently working toward the M.S. degree at Hubei University. Her main research interests include knowledge tracing and knowledge graphs.
\end{IEEEbiography}

\vspace{-25pt}

\begin{IEEEbiography}[{\includegraphics[width=1in, height=1.25in, clip, keepaspectratio]{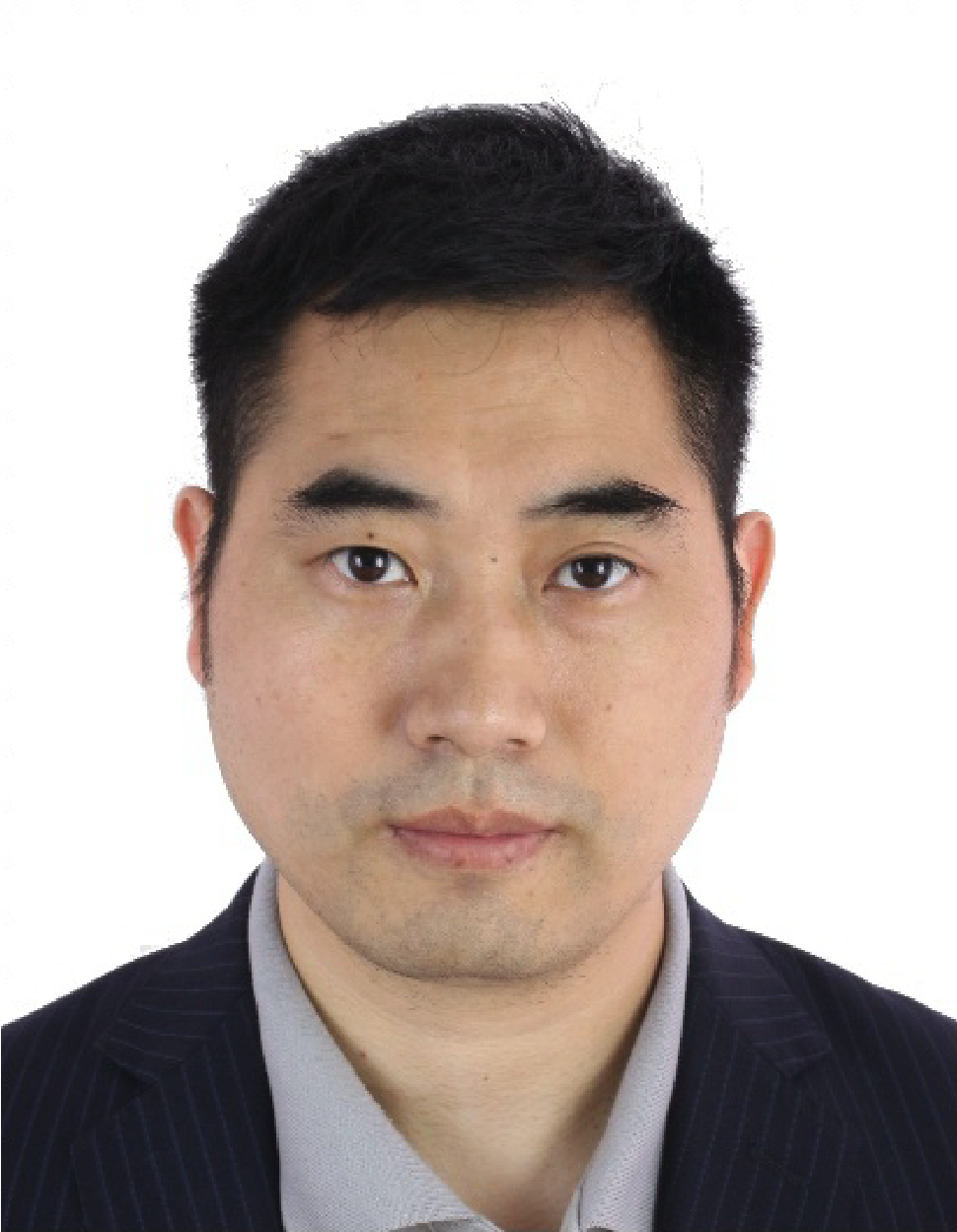}}] {Kui Xiao} received his Ph.D. degree in Wuhan University and is currently a professor in the School of Computer Science at Hubei University. He holds a senior member position in the China Computer Federation (CCF).  His current research interests are artificial intelligence technology and educational data mining.
\end{IEEEbiography}

\vspace{-25pt}

\begin{IEEEbiography}[{\includegraphics[width=1in, height=1.25in, clip, keepaspectratio]{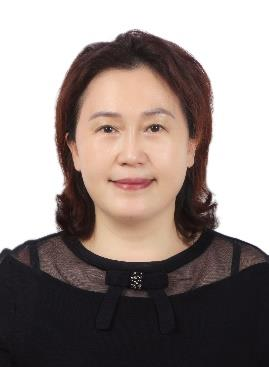}}]{Xiaoju Hou} received the Ph.D. degree from the Faculty of Artificial Intelligence in Education, Central China Normal University. She is currently an Associate Researcher and serves as the Director of the Big Data Institute for Industry-Education Integration at Guangdong Light Industry Polytechnic. She previously held the position of Deputy Secretary-General of the China Association of Educational Technology. Her research interests include educational informatization and artificial intelligence.
\end{IEEEbiography}

\vspace{-25pt}

\begin{IEEEbiography}[{\includegraphics[width=1in, height=1.25in, clip, keepaspectratio]{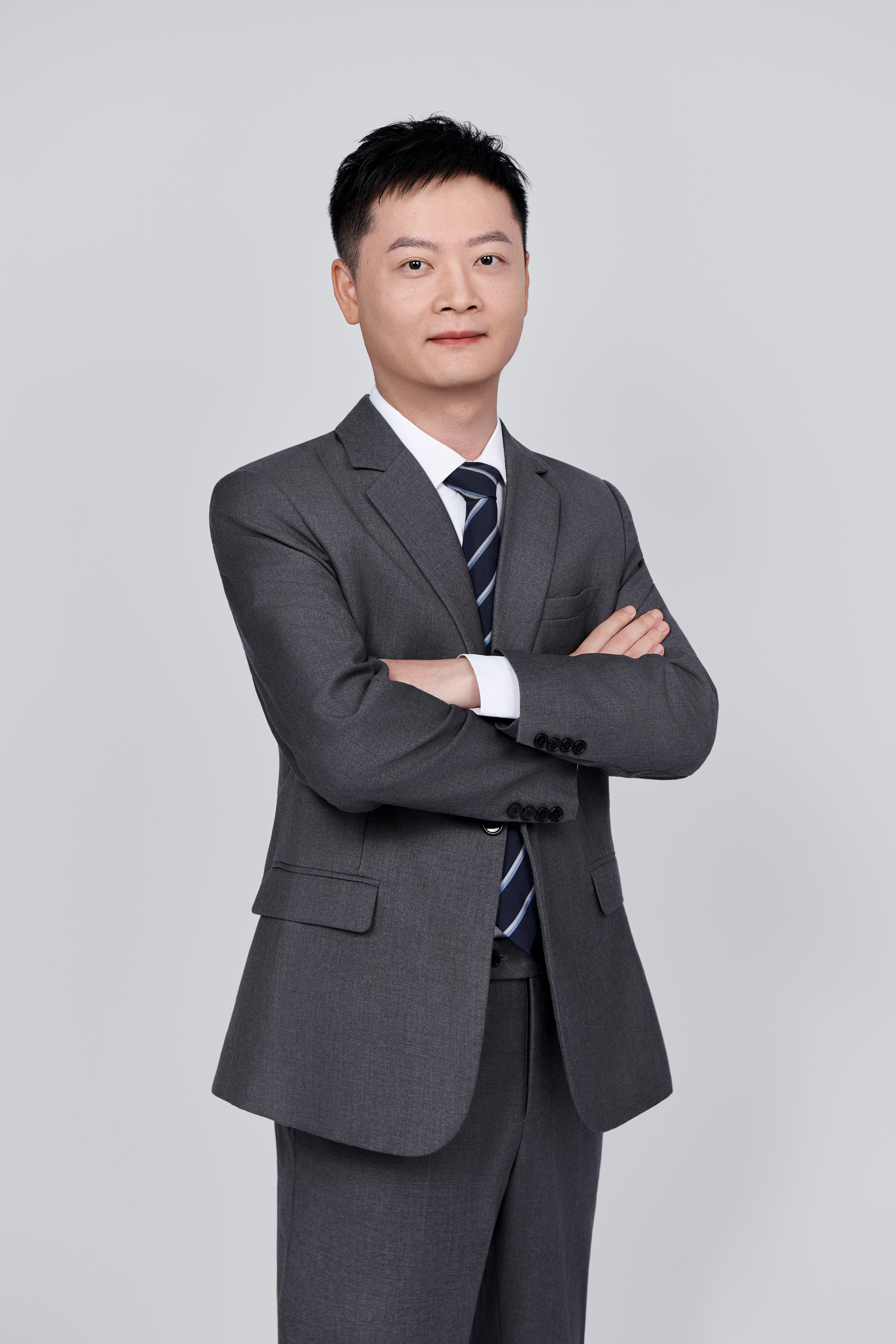}}] {Zhifei Li} received M.S. and Ph.D. degrees from the National Engineering Research Center for E-Learning at Central China Normal University in 2018 and 2021, respectively. Since October 2023, he has been an Associate Professor at the School of Computer Science, Hubei University. He has authored over thirty peer-reviewed papers in international journals and conferences such as TKDE, TNNLS, TMM, TKDD, and AAAI, with three papers selected as ESI highly cited papers. Additionally, he frequently reviews for international journals and conferences, including TKDE, TNNLS, TKDD, and AAAI. His current research interests include knowledge graphs and recommendation systems.
\end{IEEEbiography}

\end{document}